\documentclass[
    aps,
    pra,
    longbibliography,
    superscriptaddress,
    amsmath,amssymb,amsfonts, 
    tightenlines,
    twocolumn,
    ]{revtex4-2}

\usepackage[utf8]{inputenc}  
\usepackage[T1]{fontenc}     
\usepackage[english]{babel}  
\usepackage{hyperref}  
\usepackage{graphicx} 
\usepackage{float}
\usepackage[babel]{microtype}  
\usepackage{amsmath,amssymb,amsthm,bm,amsfonts,mathrsfs,bbm} 
\usepackage{setspace}
\usepackage{times}
\usepackage{braket}
\usepackage[bottom]{footmisc}
\usepackage{blochsphere}
\usepackage{pgfplots}
 
\pgfplotsset{compat=1.17}
\usetikzlibrary{calc,3d,shapes, pgfplots.external, intersections}
\usepackage{algorithm}
\usepackage{physics}
\usepackage[noend]{algorithmic}
\usepackage{eqparbox}

\usepackage{physics}
\usepackage{tikz}
\usepackage{mathtools}
\usetikzlibrary{arrows.meta}
\usepackage{csquotes}

\usepackage{xspace}  
\usepackage{pgfplots}
\usepackage{verbatim}
\usepackage{amsmath}

\def\U{{\hat U}}

\newcommand\id{\leavevmode\hbox{\small1\kern-3.3pt\normalsize1}}

\usepackage{pdfpages}
\usepackage{pgffor}
\makeatletter
\AtBeginDocument{\let\LS@rot\@undefined}
\makeatother

\begin{document}

\title[Tensor network gaussian boson sampling]{Simulating lossy Gaussian boson sampling with matrix product operators}

\author{Minzhao Liu}
\affiliation{Department of Physics, The University of Chicago, Chicago, IL 60637, USA}
\affiliation{Computational Science Division, Argonne National Laboratory, Lemont, IL 60439, USA}

\author{Changhun Oh}
\affiliation{Pritzker School of Molecular Engineering, The University of Chicago, Chicago, IL 60637, USA}

\author{Junyu Liu}
\affiliation{Pritzker School of Molecular Engineering, The University of Chicago, Chicago, IL 60637, USA}
\affiliation{Department of Computer Science, The University of Chicago, Chicago, IL 60637, USA}
\affiliation{Chicago Quantum Exchange, Chicago, IL 60637, USA}
\affiliation{Kadanoff Center for Theoretical Physics, The University of Chicago, Chicago, IL 60637, USA}
\affiliation{qBraid Co., Chicago, IL 60615, USA}
\affiliation{SeQure, Chicago, IL 60615, USA}

\author{Liang Jiang}

\affiliation{Pritzker School of Molecular Engineering, The University of Chicago, Chicago, IL 60637, USA}
\affiliation{Chicago Quantum Exchange, Chicago, IL 60637, USA}

\author{Yuri Alexeev}
\affiliation{Computational Science Division, Argonne National Laboratory, Lemont, IL 60439, USA}
\affiliation{Department of Computer Science, The University of Chicago, Chicago, IL 60637, USA}
\affiliation{Chicago Quantum Exchange, Chicago, IL 60637, USA}

\maketitle

{\bf Gaussian boson sampling, a computational model that is widely believed to admit quantum supremacy, has already been experimentally demonstrated and is claimed to surpass the classical simulation capabilities of even the most powerful supercomputers today. However, whether the current approach limited by photon loss and noise in such experiments prescribes a scalable path to quantum advantage is an open question. To understand the effect of photon loss on the scalability of Gaussian boson sampling, we analytically derive the asymptotic operator entanglement entropy scaling, which relates to the simulation complexity. As a result, we observe that efficient tensor network simulations are likely possible under the $N_\text{out}\propto\sqrt{N}$ scaling of the number of surviving photons orange$N_\text{out}$ in the number of input photons $N$. We numerically verify this result using a tensor network algorithm with $U(1)$ symmetry, and overcome previous challenges due to the large local Hilbert space dimensions in Gaussian boson sampling with hardware acceleration. Additionally, we observe that increasing the photon number through larger squeezing does not increase the entanglement entropy significantly. Finally, we numerically find the bond dimension necessary for fixed accuracy simulations, providing more direct evidence for the complexity of tensor networks.
}

\maketitle

\section{Introduction}

Exact classical simulations of quantum systems are intractable due to the exponential size of the Hilbert space. As a result, computations using quantum systems have been proposed to achieve improvement in algorithmic complexities in tasks such as integer factoring \cite{shor1994algorithms}, unstructured search \cite{grover1996a}, linear algebra \cite{harrow2009quantum}, Hamiltonian simulations \cite{berry2007Efficient,berry2014exponential,childs2010on,low2017optimal}, and more \cite{Alexeev2021}. Present-day quantum computational devices, however, are susceptible to noise and cannot be perfectly controlled. As a result, numerous approaches based on sampling outputs of randomly configured devices have been proposed to demonstrate quantum supremacy, which are especially appealing considering near-term constraints. For example, boson sampling \cite{aaronson2011computational}, a process of sampling the photon output patterns from interferometers, has resulted in numerous experimental demonstrations of quantum supremacy \cite{broome2013photonic,spring2013boson,tillmann2013experimental,crespi2013integrated,spagnolo2014experimental,carolan2014experimental,carolan2015universal,bentivegna2015experimental,zhong201812,zhong2019experimental,paesani2019generation,he2017time,loredo2017boson,wang2017high,wang2018toward,wang2019boson,zhong2020quantum,zhong2021phase,madsen2022quantum}.

However, experimental imperfections such as photon loss can have implications on the computational complexity. The effects of noise are already examined in various contexts of quantum computing experiments. For qubits cases, it has long been known that without error correction, a quantum state after a large depth with a constant level of depolarizing noise becomes very close to the maximally mixed state \cite{aharonov1996limitations}, which enables an efficient approximate simulation. One proxy of classical simulation complexity is the entanglement entropy (EE). For pure state simulations, the computational cost using tensor networks is exponential in the EE of the quantum system, which implies that systems with logarithmic growth in the EE can be efficiently simulated. Similarly, it has been argued that the density operator EE of mixed states implies a similar computational cost, albeit some nuances \cite{noh2020efficient, zhang2022entanglement, oh2021classical}. In the context of noisy random circuit sampling (RCS) \cite{arute2019quantum}, it was numerically shown that the density operator EE decreases if the circuit depth is too high for 1D \cite{noh2020efficient} and 2D \cite{zhang2022entanglement} systems. Further, the maximum achievable EE follows area law scaling, suggesting the possibility of efficient tensor network simulation. More recently, polynomial time simulation of RCS with constant depolarizing noise per gate is proven to be possible in an asymptotic regime for larger than logarithmic depths, denying the scalability of RCS \cite{aharonov2022polynomial}.

Meanwhile, in the context of boson sampling, a similar study has very recently shown that for a particular noise, which may not be experimentally relevant, there is an efficient classical algorithm for noisy boson sampling in an asymptotic regime \cite{oh2023classical}. Thus, the experimental noise might prohibit scalable quantum advantage in boson sampling, much like in RCS.
However, it still remains possible for noisy boson sampling to be scalable under realistic noises, such as photon loss \cite{oszmaniec2018classical,garcia2019simulating,renema2018classical,qi2020regimes} and partial distinguishability \cite{tichy2015sampling, renema2018efficient, renema2018classical, shchesnovich2019noise, moylett2019classically}.
Notably, the effects of photon loss are investigated in several studies. For single-photon \cite{oszmaniec2018classical, garcia2019simulating} and Gaussian boson sampling \cite{qi2020regimes, cifuentes2023classical}, when the number of photons $N_\text{out}$ surviving before measurement scales as the square root of the number of input photons $N$ ($N_{\text{out}}\propto \sqrt{N}$), classical state approximation of the output state provides an efficient method of simulation. However, the approximation error of these methods is fixed for given parameters and cannot be controlled with more resources. As a result, quantum supremacy intermediate-size experiments, where transmission is not as low as these approximate algorithms require, elude these methods.

Tensor network methods, on the other hand, allow us to control the simulation error by tuning time and memory resources, and have been used to numerically show the logarithmic scaling of the operator EE when $N_{\text{out}}\propto \sqrt{N}$ in single photon boson sampling (SPBS) \cite{oh2021classical}. However, the probabilistic nature of single-photon generation renders SPBS unscalable, and the community has long moved onto other photon sources \cite{lund2014boson,barkhofen2017driven,chakmakhchyan2017boson}. The most promising approach is Gaussian boson sampling (GBS), where no post-selection is necessary and classical simulation is hard unless some plausible complexity-theoretic conjectures are false \cite{hamilton2017gaussian,kruse2019detailed}. This allowed recent experimental demonstrations of GBS to claim quantum supremacy \cite{zhong2021phase,madsen2022quantum,deng2023gaussian}.

In this work, we investigate the operator EE scaling of GBS, which is more experimentally relevant. We show analytically that in the asymptotic limit of large $N$, the logarithmic operator EE scaling holds for $N_{\text{out}}\propto \sqrt{N}$. For numerical verification, simulation of GBS is especially difficult with tensor networks due to the infinite-dimensional local Hilbert space for each squeezed mode, which remains high even under suitable truncation and leads to dramatically increased computational cost. As a result, we develop a hardware-accelerated, supercomputing tensor network algorithm that exploits $U(1)$ symmetry, allowing us to simulate previously intractable systems such as GBS \cite{liu2023supercomputing}. We numerically verify the operator EE scaling of GBS under various loss conditions against the asymptotic estimates, and further observe that increasing the photon number through higher squeezing has little impact on EE. Finally, we explicitly calculate the bond dimension and the computational cost as the most direct evidence on the complexity. Overall, our work suggests that boson sampling with loss higher than the aforementioned scaling may be efficiently simulated with tensor networks as the system size grows.

\subsection{Related work}

The first class of methods for simulating Gaussian boson sampling are exact methods, which directly computes the loop Hafnians to determine the probability amplitudes of detection events \cite{quesada2020exact, bulmer2022boundary, quesada2022quadratic}. These methods have exponential time or space complexity, and do not deal with lossy states.

Besides exact simulation algorithms, approximate methods such as tensor network methods have been developed to reduce the simulation costs. Another approximate algorithm uses polynomial approximation of the marginals of the outputs, and the complexity is exponential in the order of approximation $k$ \cite{villalonga2021efficient}. There is no explicit use of the lossy nature of GBS. The relation between the required order $k$ and photon loss is not well understood, and therefore the time complexity scaling with loss is similarly unknown.

The most relevant class of approximate algorithms that explicitly exploits photon loss is the aforementioned classical state approximation approach \cite{qi2020regimes, cifuentes2023classical}. These methods find a classical state that approximate the squeeze states as closely as possible, and the interference outcomes of these classical states is efficient to simulate. However, this means that once the classical description is fixed, there is no control to further decrease the error by any means. It is shown that these methods can give asymptotically small error when $N_{\text{out}}\propto \sqrt{N}$ or less. The complexity is always polynomial regardless the loss scaling, but the error becomes unacceptable for higher loss. Although an earlier experiment \cite{zhong2020quantum} has been found to be potentially well described by the classical states, more recent experiments cannot be simulated with classical states. The fact that finite size quantum supremacy experiments have low loss makes this approach unsuitable.

Overall, algorithms applicable to finite size system simulations do not directly use the lossy nature of GBS, and cannot control the simulation error with full freedom. Tensor network methods give a direct measure of the simulation complexity in terms of the bond dimension. This is the precise reason why we vary experimental parameters such as loss, squeezing, and system sizes while keeping the approximation error $1-\text{Tr}(\rho)$ fixed. This unique ability of fixing the error by varying the bond dimension allows us to measure the simulation cost as a function of the experimental parameters. Further, while other sampling methods only seek to spoof the sampling benchmarks such as XEB scores and low order marginals, tensor networks directly approximate the quantum state, and offers a much richer set of information that can be potentially investigated for theoretical interest \cite{footnote}.

\section{Method}

Lossless and lossy quantum states in boson sampling can be represented by matrix product states (MPSs) and matrix product operators (MPOs) \cite{oh2021classical, huang2019simulating}. More explicitly, a general $M$-body pure state with local Hilbert space dimension $d$ can be written as
\begin{equation}
    \vert\Psi\rangle=\sum_{i_1,\dots,i_M=0}^{d-1}c_{i_1,\dots,i_M}\vert i_1,\dots,i_M\rangle,
\end{equation}
where the tensor $c_{i_1,\dots,i_M}$ fully characterizes the state $\vert \Psi\rangle$. However, tensor $c$ is $M$ dimensional, leading to $d^M$ entries in storage. To reduce the storage cost, the standard MPS ansatz represents the state in a compressed manner as
\begin{align}\label{MPS}
c_{i_1,\dots,i_M}=\sum_{\alpha_0,\dots,\alpha_M=0}^{\chi-1} & \Gamma_{\alpha_0\alpha1}^{[1]i_1}\lambda_{\alpha_1}^{[1]}\Gamma_{\alpha_1\alpha2}^{[2]i_2}\lambda_{\alpha_2}^{[2]}\times\\\nonumber
& \dots\lambda_{\alpha_{M-1}}^{[M-1]}\Gamma_{\alpha_{M-1}\alpha_M}^{[M]i_M},
\end{align}
where $\chi$ is the so-called bond dimension, and larger $\chi$ corresponds to a lower approximation error. In the MPS representation, each $\Gamma^{[j]}$ tensor contains information about the $j$-th body, and $\Gamma^{[j]i_j}_{\alpha_{j-1}\alpha_j}$ captures its amplitudes in state $i_j$, conditioned on the states of the left and right neighbors specified by the $\alpha_{j-1}$ and $\alpha_j$ incides.  The $\lambda$ tensors can be understood as singular values of Schmidt decomposition, which we discuss in more detail in E.q. \ref{schmidt}. The MPS represents this large tensor as a contraction (sum over the dummy or virtual $\alpha$ indices which capture correlations between particles) of a chain of tensors. One can observe that the $i$ indices representing the physical degrees of freedom remain open (unsummed). The memory complexity of the MPS is $O(\chi^2dM)$, and $\chi$ can be adjusted to represent $c$ with the desired accuracy. Further, one can efficiently perform local unitary operations on the MPS and calculate expectation values of local observables with complexity $O(d^4\chi^3)$.

Lossy boson sampling can be simulated using an MPO, which is essentially an MPS with additional dual indices. Time evolving the MPO can be accomplished by contracting it with Kraus operators. Further, in the case where loss is uniform (equal photon loss at every beam splitter), all losses can be moved to the initial state since loss commutes with linear optical transformations. As a result, full Kraus operator-based simulation of noisy channels is not necessary. We can represent the initial lossy state using an MPO, vectorize the MPO, build a tensor network analogous to E.q. \ref{MPS}, and update the MPO by contracting it with unitaries \cite{huang2019simulating,singh2011tensor,guo2019matrix}. Specifically, vectorization of a density operator is defined as
\begin{align}
&\hat{\rho}=\sum_{k,k'}\vert k\rangle\rho_{k,k'}\langle k'\vert\nonumber\\
\rightarrow & \vert\hat{\rho}\rangle\rangle=\sum_{k,k'}\rho_{k,k'}\vert k,k'\rangle=\sum_K\rho_K\vert K\rangle,
\end{align}
where $K$ is the combined index for $k, k'$. Applications of the unitary is also slightly modified since vectorization of $U\hat{\rho}\U^\dag$ becomes
\begin{equation}
    \sum_{k,k'}U_{j,k}\rho_{k,k'}U^\dag_{k',j'}\rightarrow\sum_K \mathcal{U}_{J,K} \rho_K,
\end{equation}
where $\mathcal{U}_{J,K}=U_{j,k}U^*_{j',k'}$.

For lossless, pure state systems with large local dimensions $d$, the $O(d^4\chi^3)$ computational complexity becomes significant. For lossy, mixed state simulations with MPOs, the complexity becomes $O(d^8\chi^3)$, which is especially problematic for systems with large $d$ such as GBS. Fortunately, for systems with global symmetry, such as particle number or total spin conservation, another level of reduction is possible. Symmetry preserving operators can be expressed as a direct sum $\hat{T}=\bigoplus_n \hat{T}_n$ where $\hat{T}_n$ preserves the subspace $\mathbb{V}_n$ corresponding to some conserved charge $n$, and tensors can be therefore written in a block diagonal form and stored efficiently \cite{singh2011tensor}. Computations are performed on different blocks independently, and the time cost also reduces due to the non-linear polynomial complexity.

An important class of quantum systems have global $U(1)$ symmetry, which arises when the system has some kind of conserved charge \cite{singh2011tensor}. Examples of such systems include the hardcore Bose Hubbard model \cite{aizenman2004bose}, the spin-$1/2$ XXZ quantum spin chain \cite{alcaraz1989}, boson sampling \cite{aaronson2011computational}, quantum walk \cite{kitagawa2010exploring,childs2010on,childs2013universal,cai2021multiparticle,schreiber2012a}, and monitored quantum circuits \cite{agrawal2022entanglement}. A model is said to be $U(1)$ symmetric if the Hamiltonian commutes with the total charge operator \cite{singh2011tensor}
\begin{equation}
    [\hat{H},\hat{N}]=0.
\end{equation}
As a result, evolution under such Hamiltonians must preserve the charge number operator. More generally, systems can preserve a global $U(1)$ symmetry if the applied unitaries preserve the global charge.

In the case of boson sampling, the unitaries preserve the global photon number. As we discussed earlier, loss can be commuted to initialization, and time evolution is fully $U(1)$ symmetric. To exploit the $U(1)$ symmetry, one can define a so-called charge which is the number of photons to the left of the bipartition. The overall effect is that the $\Gamma$ tensors lose their $i$ indices corresponding to the physical degree of freedom (local photon number), reducing the memory complexity by a factor of $d$ (see Appendix \ref{app:u1} for more details). This is instead captured by the size $\chi$ 1-d charge tensors $c$. Second, the size of the matrices that we decompose with SVD is also reduced to at most $\chi\times\chi$ instead of $\chi d\times\chi d$. Therefore, the use of $U(1)$ symmetry significantly reduces the memory and time complexity of the algorithm.

The MPS formulation is especially convenient for quantifying entanglement. If we perform the Schmidt decomposition on the quantum state, which is to express the wavefunction as the sum of tensor products of states of two subsystems $A$ and $B$
\begin{equation}
    \vert\Psi\rangle=\sum_\alpha \lambda_{\alpha}\vert\alpha_A\rangle\vert \alpha_B\rangle,\label{schmidt}
\end{equation}
where $\{\vert\alpha\rangle\}$ forms a basis set for each subsystem, we reveal the entanglement between the two subsystems, and the entanglement entropy (EE) given by
\begin{equation}
    -\sum_\alpha\lambda_\alpha^2\log{\lambda_\alpha^2}
\end{equation}
quantifies how much entanglement there is. Conveniently, if the subsystems are bipartitions of the MPS at site $\ell$, the Schmidt decomposition singular values $\lambda_\alpha$'s would be the MPS singular values $\lambda_{\alpha_\ell}^{[\ell]}$, allowing us to compute the MPS EE. For a mixed state represented by a vectorized MPO, we can formally perform Schmidt decomposition, identify the singular values $\lambda_\alpha$ with $\lambda_{\alpha_\ell}^{[\ell]}$ of the MPO, and similarly compute the MPO EE. In both cases, higher EE means more uniformly distributed singular values, and truncation leads to a higher approximation error. Larger bond dimensions are necessary to simulate systems with larger EE to fixed accuracy.

\section{Results}

\subsection{Supercomputing $U(1)$-symmetric tensor network algorithm}

Details of the numerical protocols of time evolving the $U(1)$-symmetric tensor network are available in Appendix \ref{app:u1}. CPU based implementations have already been adopted in single photon boson sampling simulations, but the computational cost for Gaussian boson sampling is still too high. Increasing the parallelism through the use of GPU and multi-node parallel computing is necessary. A naive implementation where each GPU thread computes one tensor entry performs worse than the CPU implementation due to technical reasons discussed in Appendix \ref{app:u1}. However, with implementation innovations in this work, we achieve significant run time reduction. Table \textrm{I} shows the simulation time in seconds of different implementations for a lossy boson sampling experiment with 12 modes, 10 input squeezed modes, bond dimension 1024 and 8192, photon loss rate 0.55, and local Hilbert space dimension 15. Overall, the fully parallel implementation on six nodes (4 NVIDIA A100 GPUs each) is on the order of a 1000 times faster than the 32-core CPU implementation.

\begin{table}
\centering
\begin{tabular}{ |c|c|c|c|c| }
 \hline
 & CPU & single-GPU & One node & Six nodes \\ 
 \hline
$\chi=1024$ & 7966 & 126 & 60 & 42 \\
 \hline
 $\chi=8192$ & >259000 & 2066 & 1045 & 322 \\
 \hline
\end{tabular}
\caption{Simulation time in seconds.}
\end{table}

\subsection{Analytic asymptotic entanglement entropy scaling}

We provide the asymptotic MPO EE scaling under various loss conditions. Specifically, we consider cases where the number of photons surviving before measurement $N_\text{out}$ scales with the number of input photons $N$, with different scaling exponents $0<\gamma\leq 1$ ($N_\text{out}=\beta N^\gamma$ and the transmission rate is $\mu=\beta N^\gamma/N$, where $\beta$ is a constant). Specifically, $\gamma=1$ corresponds to a constant photon loss rate as the system size grows, whereas $0<\gamma<1$ corresponds to increasing loss rates as the system size grows. This is a reasonable scaling because loss should increase with the system size due to various experimental limitations such as an increase in the depth of beam splitters required to obtain a sufficiently Haar random state.

After derivations shown in Appendix \ref{proof}, we obtain the following scaling for the MPO EE:
\begin{align}
    S_1(\vert \hat{\rho}\rangle\rangle)&=O\left(N\left(\frac{\beta N^\gamma}{N}\right)^2\log_2{\left(\frac{\beta N^\gamma}{N}\right)}\right)\nonumber\\
    &=O(N^{2\gamma-1}\log_2{N}).
\end{align}
Similarly, for the R\'enyi entropy with $\alpha<1$, we have
\begin{equation}
    S_\alpha=O(N^{1-2(1-\gamma)\alpha}).
\end{equation}
For $\alpha>1$, we have
\begin{equation}
    S_\alpha=O(\frac{\alpha}{1-\alpha}N^{2\gamma-1}).
\end{equation}

The MPO EE scaling becomes logarithmic when $\gamma=1/2$. For an MPS algorithm, a logarithmic scaling of the MPS EE already rigorously implies a polynomial time complexity for the tensor network algorithm at fixed 2-norm distance between the ideal and approximate state. This implies efficient fixed fidelity simulation. The situation for the MPO algorithm is trickier. The logarithmic MPO EE now implies efficient simulation for fixed 2-norm distance between the vectorized states, which is also the 2-norm distance between the density operators. However, for fixed fidelity, one needs to bound the 1-norm distance, and the relationship $K\vert\vert A\vert\vert_2\geq\vert\vert A\vert\vert_1$, where $K$ is the dimension of the Hilbert space, means that the one norm cannot be efficiently bounded. In some cases, the MPO EE decreases as the system size increases, reducing the required bond dimension to bound the 2-norm distance, but the required bond dimension to bound the 1-norm distance may still increase. This is the case for a sufficiently low $\gamma$ such as $\gamma=\frac{1}{4}$.

\subsection{Numerical asymptotic estimates of the entanglement entropy}

We estimate the asymptotic MPO EE under photon survival scaling $N_{\text{out}}\propto N^\gamma$ with $\gamma=\frac{1}{4},\frac{1}{2},1$. To make a fair comparison against SPBS, the squeezing parameter is fixed at $r=0.88$, which averages to approximately one photon per squeezed mode. Appendix \ref{app:asymptotic} discusses how we obtain the estimates for very large system sizes where direct MPO simulations are impractical.

\begin{figure} [ht]
   \begin{center}
   \includegraphics[width=8.6cm]{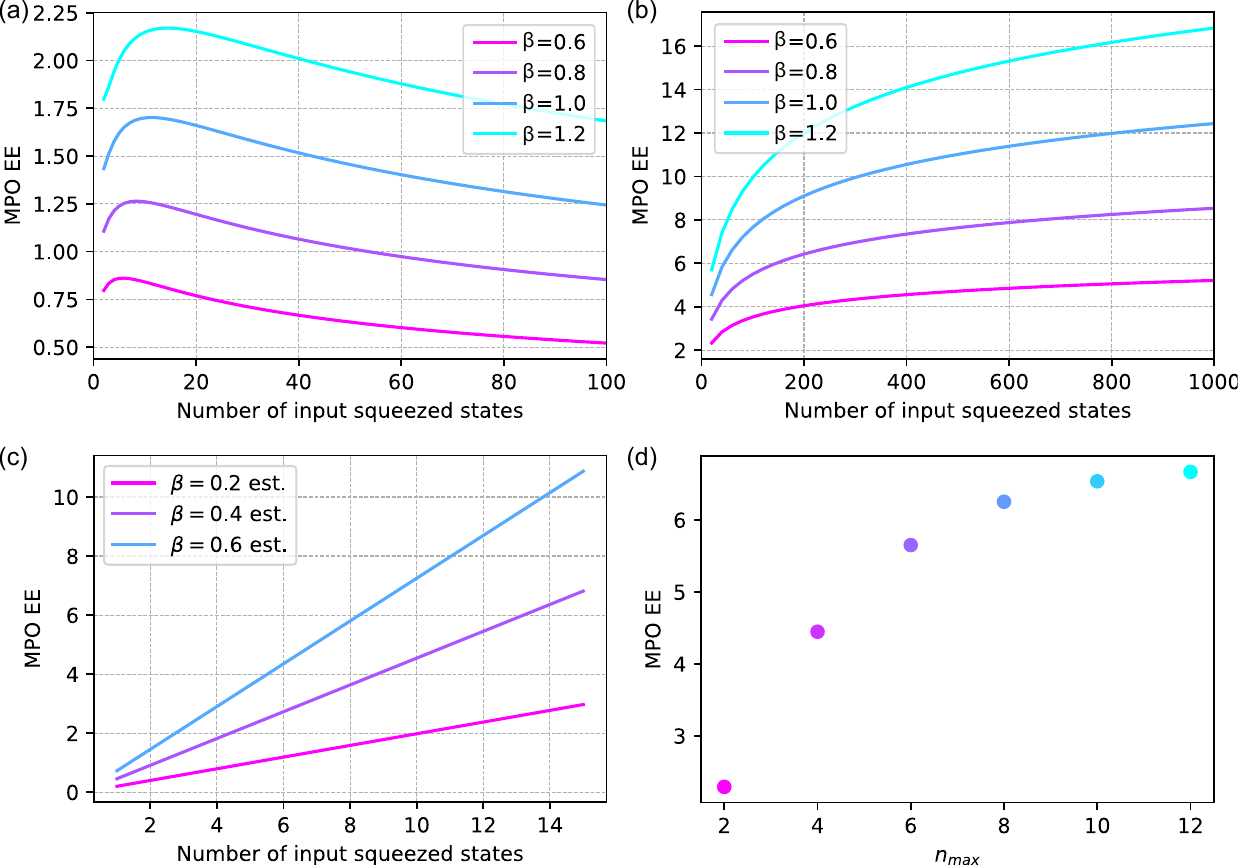}
   \end{center}
   \caption{Operator entanglement entropy vs. the number of input squeezed modes for different photon survival scaling $N_{\text{out}}=\beta N^\gamma$ at $r=0.88$. (a) $\gamma=\frac{1}{4}$. (b) $\gamma=\frac{1}{2}$. (c) $\gamma=1$. (d) Convergence of MPO EE with increasing $n_\text{max}$ for $N=50,\beta=1,\gamma=\frac{1}{2},r=0.88$.}
   { \label{asymptotic}
}
   \end{figure}

Fig. \ref{asymptotic} shows the asymptotic estimates with $n_\text{max}=8$ (maximum number of photons per density operator $\rho_j$ that is simulated, see Appendix \ref{app:asymptotic}) for large system sizes. Similar to what is observed in SPBS simulations, GBS shows MPO EE reduction when the loss is sufficiently high for $\gamma=\frac{1}{4}$, logarithmic scaling for $\gamma=\frac{1}{2}$, and linear scaling for $\gamma=1$. A similar linear increase in MPO EE with $\beta$ is also observed in all three cases. Further, we also show the numerical convergence of our asymptotic MPO EE estimates by increasing the cut-off of the initial maximum photon number $n_\text{max}$ for the squeezed states.

\begin{figure*} [ht]
   \begin{center}
   \includegraphics[width=12.9cm]{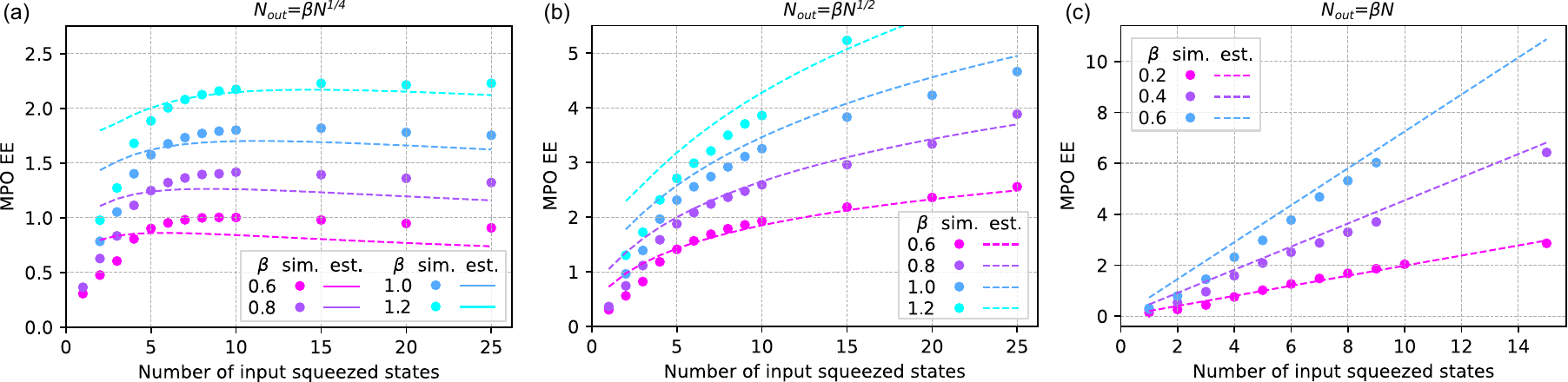}
   \end{center}
   \caption{Operator entanglement entropy vs. the number of input squeezed modes for different photon survival scaling $N_{\text{out}}=\beta N^\gamma$ at $r=0.88$. Details of experiment configurations can be found in Methods. (a) $\gamma=\frac{1}{4}$. (b) $\gamma=\frac{1}{2}$. (c) $\gamma=1$. Dots are results obtained from full simulations using $U(1)$ symmetry. Dashed lines are estimates using asymptotic assumptions.}
   { \label{EE_scaling}
}
   \end{figure*}
   
\begin{figure*} [ht]
   \begin{center}
   \includegraphics[width=12.9cm]{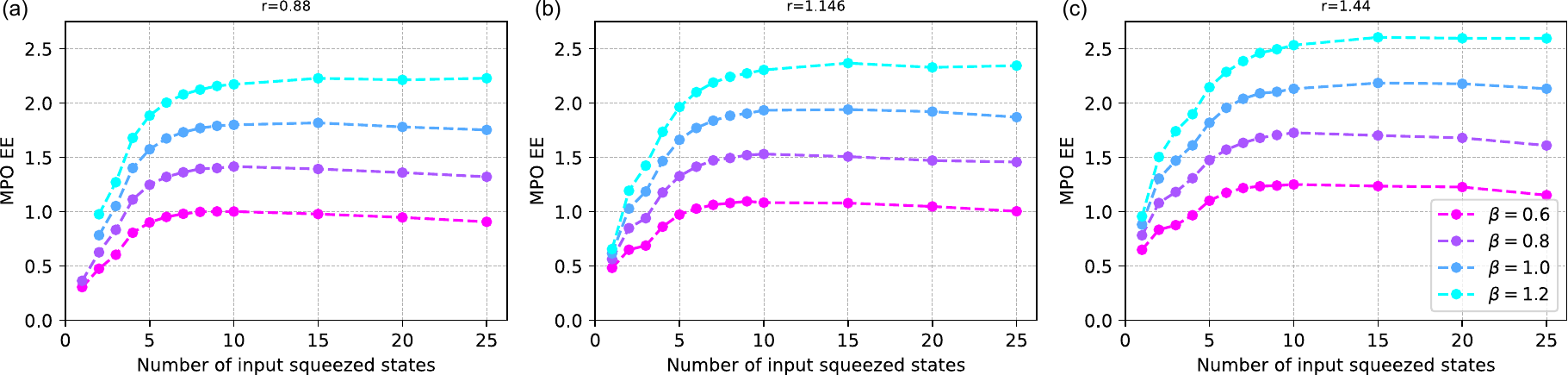}
   \end{center}
   \caption{Operator entanglement entropy vs. the number of input squeezed modes for different squeezing parameters $r$. Dashed lines are guides to the eye. (a) $r=0.88$, averaging approximately 1 photon per mode. (b) $r=1.146$, averaging approximately 2 photons per mode. (b) $r=1.44$, averaging approximately 4 photons per mode.}
   { \label{r_scaling}
}
   \end{figure*}

\subsection{Finite-size entanglement entropy from simulations}

We further conduct full MPO simulations of GBS and numerically calculate the MPO EE. The MPO EE obtained from the full simulations and asymptotic estimates agree quantitatively, as shown in Fig. \ref{EE_scaling}. However, we observe that the quality of agreement is poor when MPO EE is small such as in many $\gamma=\frac{1}{4}$ data points when the number of input squeezed states $N$ is small. In the regime of small MPO EE but large $N$, we attribute the disagreement to the formal differences between regular MPOs and MPOs in a $U(1)$ symmetric form. This is easy to see as even a product state can have non-zero $U(1)$ symmetric MPO EE simply due to the existence of different charges. For small $N$, we expect the quality of the approximation to be poor because we are no longer in the asymptotic limit. Further disagreement can also be attributed to the fact that the $U(1)$ symmetric full simulations are limited by the bond dimension. We ensure that all plotted data points are simulated to $1-\text{Tr}(\hat{\rho})<0.1$, which previous work established as a good proxy to the fidelity and the total variation error that is computationally lightweight \cite{zhang2022entanglement, oh2021classical}.

Lastly, we investigate the effect of squeezing on MPO EE with our full $U(1)$ symmetric simulations. We choose to investigate $\gamma=\frac{1}{4}$ for easier simulation. Fig. \ref{r_scaling} shows an increase in MPO EE with increasing squeezing parameter $r$. It is important to note that the average number of output photons scales with the average number of input photons $N$, not the number of squeezed states. This means that for the same number of input squeezed states and $\beta$, a higher squeezing parameter has a higher loss. Increasing the average number of photons per squeezed mode from 1 to 2 and 4 only moderately increases the MPO EE compared to increasing $N$. This observation is similar to the previous finding for Fock state boson sampling: if the number of input modes stays the same and the number of photons per mode increases, the MPO EE grows slowly and can be efficiently simulated \cite{oh2021classical}. 

\begin{figure} [ht]
   \begin{center}
   \includegraphics[width=8.6cm]{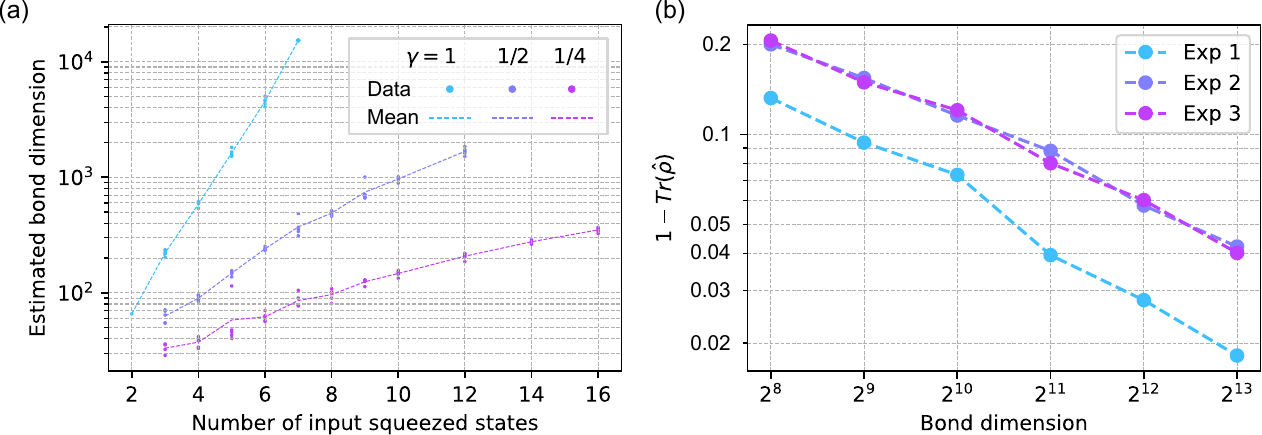}
   \end{center}
   \caption{Analysis of bond dimension, system size, and error. Details can be found in Methods. (a) Bond dimension needed to reach accuracy $1-\text{Tr}(\hat{\rho})=0.02$ vs. the number of input squeezed modes photon survival scaling $N_{\text{out}}=0.4 N^\gamma$ at $r=0.88$. Dots are individual estimates of the bond dimension obtained from full simulations using $U(1)$ symmetry. Dashed lines are the means. (b) Reduction in $1-\text{Tr}(\hat{\rho})$ error as bond dimension increases for three different experimental configurations.}
   { \label{bond_dimension}
}
   \end{figure}

Our numerical findings on the MPO EE growth for different loss scalings have complexity implications, but there is a lack of rigorous correspondence between MPO EE and simulation time. To make the statement on simulation complexity more direct, we validate the bond dimension growth explicitly. This is helpful in particular because the computational complexity is qubic in the bond dimension, both due to SVD and matrix multiplication. We show in Fig. \ref{bond_dimension} the growth of bond dimension in the system size for fixed accuracy of $1-\text{Tr}(\hat{\rho})=0.02$. Previous work has established that $1-\text{Tr}(\rho)$ is a good proxy for the fidelity \cite{zhang2022entanglement} and the total variational distance \cite{oh2021classical}, which is the gold standard benchmark for boson sampling sample quality. It is clear that constant loss leads to exponential growth in the bond dimension. In higher loss cases, growth is much more moderate and appears sub-exponential. We also validate that increasing the bond dimension efficiently reduces the simulation error. We choose three experiments and simulated them with different bond dimensions.

\section{Discussion}

We show analytically that the matrix product operator entanglement entropy of boson sampling scales logarithmically under high loss, which we numerically verify using $U(1)$ symmetric tensor networks. We also numerically observe that increasing the photon number by squeezing has little impact compared to increasing the number of input squeezed modes. The computational complexity is also directly studied by calculating the bond dimension. This extends the previous entanglement entropy results for single photon boson sampling to the more experimentally relevant Gaussian boson sampling, and extends the efficient simulation results using classical state approximation algorithms to an algorithm with controllable error, a necessary condition for simulating intermediate-size experiments with practical transmission rates. As a result, our analysis is more relevant to current quantum supremacy boson sampling experiments.

Although the MPO formalism intrinsically assumes a 1D architecture of the interferometer, the fact that we are simulating Haar random unitaries means that our findings are architecture independent. However, if we want to simulate high dimensional low depth systems that are not Haar random \cite{Deshpande2022quantum,madsen2022quantum}, one potential direction to move forward is to adopt more exotic tensor networks such as projected entanglement pair states (PEPS), and $U(1)$ symmetric forms of generic tensor networks can be constructed in principle \cite{singh2011tensor}.

\section*{Data availability}
 Data used to generate the figures are available upon request from the authors.

\section*{Code availability}
The code used to generate the data and figures is available in the GitHub repository \url{https://github.com/sss441803/BosonSampling}.

\section*{Acknowledgements}
This research used the resources of the Argonne Leadership Computing Facility, which is a U.S. Department of Energy (DOE) Office of Science User Facility supported under Contract DE-AC02-06CH11357. Y.A. acknowledges support from the Office of Science, U.S. Department of Energy, under contract DE-AC02-06CH11357 at Argonne National Laboratory and Defense Advanced Research Projects Agency (DARPA) under Contract No. HR001120C0068. L.~J.~acknowledges support from the the ARO(W911NF-23-1-0077), ARO MURI (W911NF-21-1-0325), AFOSR MURI (FA9550-19-1-0399, FA9550-21-1-0209), AFRL (FA8649-21-P-0781), DoE Q-NEXT, NSF (OMA-1936118, ERC-1941583, OMA-2137642), NTT Research, and the Packard Foundation (2020-71479). J.L. is supported in part by International Business Machines (IBM) Quantum through the Chicago Quantum Exchange, and the Pritzker School of Molecular Engineering at the University of Chicago through AFOSR MURI (FA9550-21-1-0209). M.L. acknowledges support from DOE Q-NEXT. C.O. acknowledges support from the ARO (Grants
No. W911NF-18-1-0020 and No. W911NF-18-1-0212), ARO
MURI (Grant No. W911NF-16-1-0349), AFOSR MURI
(Grants No. FA9550-19-1-0399 and No. FA9550-21-1-0209), DOE Q-NEXT, NSF (Grants No. EFMA-1640959, No. OMA-1936118, No. EEC-1941583), NTT Research, and the Packard
Foundation (2013-39273).

\section*{Competing Interests}
The authors declare no competing financial or non-financial interests.

\section*{Author Contributions}
M.L. developed the majority of the software, performed all numerical simulations and theoretical derivations, and wrote the majority of the manuscript. C.O. provided the original CPU implementation, developed the MPO initialization scheme, performed preliminary theoretical work, and contributed significantly to manuscript editing. All other authors contributed novel ideas, provided numerous scientific and writing improvements about the paper, and participated in discussions that shaped the project in a substantial manner and the understanding of its broader impact.

\appendix

\section{Details of numerical experiments}

We carry out our GPU numerical simulations using the Polaris system at the Argonne Leadership Computing Facility (ALCF). Each node has a single 2.8 GHz AMD EPYC Milan 7543P 32-core CPU with 512 GB of DDR4 RAM and four Nvidia A100 GPUs connected via NVLink. We perform our CPU numerical simulations using the Bebop system at ALCF. Each node has a single 2.10 GHz Intel Xeon E5-2695v4 32-core CPU with 128GB of DDR4 RAM.

All full $U(1)$ symmetric MPO simulations have $M=max(20,4N)$ modes, and the local Hilbert space dimension $d$ is chosen such that $<\%1$ of the probability is truncated. Because the left-most charge is the sum of all photons, the required $d$ is higher for higher squeezing parameters $r$ and higher numbers of input squeezed states. Global Haar random interferometers are used and constructed using an $M$-layer array \cite{Russell2017direct}.

For Fig. \ref{EE_scaling} and \ref{r_scaling}, we simulate the system until there is no MPO EE increase for at least 10 layers. This is reasonable since we only care about the maximum MPO EE throughout simulation which captures the computational cost, and the MPO EE generically increases as more depths are simulated until saturation, after which the MPO EE decreases slowly. All data points of MPO EE are obtained by only a single experiment, as we observe that no significant noise is present in our results because we only extract the maximum. We ensure that all plotted data points are simulated to $1-\text{Tr}(\hat{\rho})<0.1$, which previous work established as a good proxy to the total variation error that is computationally lightweight. The largest simulation is for linear scaling simulations with $d=18,\beta=0.4,N=15,M=60,\chi=16384$ ran on 10 Polaris nodes with 40 GPUs in total.

For Fig. \ref{bond_dimension}, the full depth of the interferometer is simulated. This is because simulating each layer produces additional error, and therefore affects the required bond dimension. For each configuration in Fig. \ref{bond_dimension} a, the bond dimension starts with a small value and doubles if the error exceeds $0.02$. Once the bond dimension is large enough to exceed the desired accuracy, the bond dimension is refined a few more times to obtain a more precise estimate. For linear scaling and $7$ input squeezed states, the simulation is expensive and we verified that setting $\chi=15296$ produced $1-\text{Tr}(\hat{\rho})=0.020,0.022,0.021,0.021,0.019$ across $5$ different experiments and used $15296$ as the estimated bond dimension. The three experiments for Fig. \ref{bond_dimension} b are $M=10,15,20$ and $\mu=0.2,0.15,0.1$ respectively, and each data point is obtained from one simulation.

\section{Proof of asymptotic entanglement entropy scaling}\label{proof}

In this section, we discuss the derivation of the operator EE scaling. We consider boson sampling where $N$ independent input optical modes are sent into a linear optical interferometer. The interferometer has $M$ modes, which can be larger than $N$, making $M-N$ modes at the input vacuum states. As photons interact throughout the interferometer, the quantum state gets transformed according to a unitary matrix describing the interferometer, and photons eventually exit the $M$ optical modes with non-trivial correlation. For boson sampling, the claim is that this process is hard to simulate for a sufficiently random unitary describing the interferometer.

Formally, the quantum state of $N$ independent and identical modes can be written as:
\begin{equation}
\vert \psi_{\text{in}}\rangle=\otimes^N_{j=1}\vert\psi\rangle_j=\otimes^N_{j=1}\left(\sum_{n=0}^\infty c_n\frac{\hat{a}_j^{\dag n}}{\sqrt{n!}}\right)\vert 0\rangle,
\end{equation}
where $\hat{a}_j^{\dag}$ is the input creation operator for mode $j$. The corresponding density operator is
\begin{equation}\label{single_in}
    \hat{\rho}_\text{in}=\otimes_{j=1}^N\vert \psi\rangle_j\langle\psi\vert_j=\otimes_{j=1}^N\hat{\rho}_{j,\text{in}},
\end{equation}
where the single input mode density operator is
\begin{equation}
    \hat{\rho}_{j,\text{in}}=\sum_{n,m=0}^\infty c_{n}c_{m}^* \vert n\rangle_j\langle m\vert_j.
\end{equation}

The action of an $M$-mode beam splitter array is to transform input creation operators:
\begin{equation}
\hat{a}_j^{\dag}\rightarrow \hat{b}_j^\dag=\sum_{k=1}^M U_{jk}\hat{a}_{k}^{\dag},
\end{equation}
where $\hat{b}^\dag_j$ is the output creation operator for mode $j$. To study the entanglement entropy between bipartitions separated at the $l$-th mode, we can define the normalized up and down bipartition creation operators $\hat{B}^\dag_{\text{u},j},\hat{B}^\dag_{\text{d},j}$ as
\begin{equation}
    \cos{\theta_j}\hat{B}_{\text{u},j}^{\dag}=\sum_{k=1}^l U_{jk}\hat{a}_{k}^{\dag},\quad 
    \sin{\theta_j}\hat{B}_{\text{d},j}^{\dag}=\sum_{k=l+1}^M U_{jk}\hat{a}_{k}^{\dag},
\end{equation}
with normalizations
\begin{equation}\label{normalization}
    \cos^2{\theta_j}=\sum_{k=1}^{l}\vert U_{jk}\vert^2,\quad
    \sin^2{\theta_j}=\sum_{k=l+1}^{M}\vert U_{jk}\vert^2.
\end{equation}
In the collision-free cases where $M\geq N^2$, the bipartition creation operators satisfy the canonical commutation relations
\begin{eqnarray}
[\hat{B}_{\text{u},j},\hat{B}_{\text{u},k}^{\dag}]=\delta_{jk},\quad
[\hat{B}_{\text{d},j},\hat{B}_{\text{d},k}^{\dag}]=\delta_{jk},\\\nonumber
[\hat{B}_{\text{u},j},\hat{B}_{\text{d},k}]=0,\quad
[\hat{B}_{\text{u},j},\hat{B}_{\text{d},k}^{\dag}]=0.
\end{eqnarray}
As a result, one can define the mutually orthogonal bipartition number states
\begin{eqnarray}
    \hat{B}_{\text{side},j}^{\dag k}\vert 0\rangle=\frac{\vert k\rangle_{\text{side},j}}{\sqrt{k!}},\text{side}\in\{\text{u,d}\}.
\end{eqnarray}

The above formalism, described in \cite{oh2021classical}, allows us to calculate the MPO EE without explicitly constructing the output state given the unitary representing the interferometer. Specifically, the details of the unitary matrix are hidden in the $\vert k\rangle_{\text{side},j}$ states constructed to satisfy orthogonality.

In this picture, the action of the unitary is to transform the input basis in the following way:
\begin{equation}\label{transform}
    \vert n\rangle_j\rightarrow\sum_{k_j=0}^n\sqrt{\binom{n}{k_j}}\cos^{k_j}{\theta_j}\sin^{n-k_j}{\theta_j}\vert k_j\rangle_{\text{u},j}\vert n-k_j\rangle_{\text{d},j},
\end{equation}
and therefore (omitting the $j$ index),
\begin{equation}\label{kun}
\langle k_\text{u},k_\text{d}\vert U\vert n\rangle=\sqrt{\binom{n}{k_\text{u}}}\cos^{k_\text{u}}{\theta}\sin^{k_\text{d}}{\theta}\delta(k_\text{u}+k_\text{d}-n).
\end{equation}

If we apply this basis transform due to the unitary to the input lossless density operator $\hat{\rho}_\text{int}$, each single mode input density operator $\hat{\rho}_{j,\text{in}}$ in Eq. \ref{single_in} would transform independently. The full density operator remains a product of input modes in the new basis, and we have 
\begin{equation}
\hat{\rho}_\text{out}=\otimes_j^N\hat{\rho}_{j,\text{out}}.
\end{equation}
Although each $\hat{\rho}_{j,\text{out}}$ can be identified with an input mode $j$, $\hat{\rho}_{j,\text{out}}$ is no longer a single mode state, and is instead supported over all modes. Each $\hat{\rho}_{j,\text{out}}$ has some EE because it describes a state over both partitions, and the full system EE is additive in $j$ due to the tensor product structure of $\hat{\rho}_\text{out}$. Therefore, the system EE scales linearly with the number of input modes $N$, and classical simulation of lossless boson sampling is always inefficient in $N$.

We can extend this analysis to lossy cases. Assuming that loss is uniform throughout the interferometer, loss commutes with all linear optical transforms and can be applied to the initial pure state. The basis transform \ref{transform} due to the unitary is still independent on $j$, and the total output density operator is still in a product form with
\begin{equation}
\hat{\rho}_{j,\text{out}}=U\mathscr{E}_\text{loss}(\hat{\rho}_{j,\text{in}})U^\dag.
\end{equation}
Therefore, the linear scaling of EE in $N$ remains, and MPO simulations of lossy boson sampling is also inefficient in $N$.

However, as the number of input modes $N$ increases, the complexity of the interferometer must grow as well in order to maintain reasonable randomness in the interferometer unitary and hardness of classical simulation. As a result, the depth of the interferometer should scale with the $N$, which leads to scaling of the transmission rate $\mu$ in $N$. The entanglement entropy for each $\hat{\rho}_{j,\text{out}}$ decreases as $N$ increases, leading to an overall entanglement entropy that grows sublinearly, potentially allowing efficient simulation. To understand the scaling in loss and transmission, consider the Kraus operators corresponding to the single input state photon loss channel in the limit of small $\mu$ (from now on we ignore the mode index $j$)
\begin{align}
\hat{\rho}_\text{lossy}&=\mathscr{E}_\text{loss}(\hat{\rho}_\text{in})=\sum_{n_{\text{loss}}=0}^{n_{\text{max}}}K^{(n_{\text{loss}})}\hat{\rho}_\text{in} K^{(n_{\text{loss}})\dag},\\
K^{(n_{\text{loss}})}&=\sum_{n_{\text{out}},n_{\text{in}}=0}^{n_{\text{max}}}K^{(n_{\text{loss}})}_{n_{\text{out}},n_{\text{in}}} \vert n_{\text{out}} \rangle \langle n_{\text{in}}\vert\\
K^{(n_{\text{loss}})}_{n_{\text{out}},n_{\text{in}}}&=\begin{cases}
       \sqrt{\binom{n_{\text{in}}}{n_{\text{out}}}\mu^{n_\text{out}}(1-\mu)^{n_\text{loss}}} &\quad\text{if }n_{\text{in}}-n_{\text{out}}=n_{\text{loss}}\\
      0 &\quad\text{otherwise},
     \end{cases}
\end{align}
where $K^{(n_\text{loss})}\in \mathbb{C}^{n_{\text{max}}+1,n_{\text{max}}+1}$ captures processes that lose $n_{\text{loss}}$ photons, and we limit the maximum photon number to $n_{\text{max}}$. The lossy density operator can be given in the input $\vert n\rangle$ basis in index notation:
\begin{align}\label{lossy}
\rho_{\text{lossy} ~m,n}&=\sum_{n_{\text{loss}}=0}^{n_{\text{max}}}\sum_{k,l}K^{(n_{\text{loss}})}_{m,k}\rho_{\text{in}~k,l}K^{(n_{\text{loss}})\dag}_{l,n}\nonumber\\
&=\sum_{n_{\text{loss}}=0}^{n_{\text{max}}}O(\mu^\frac{m}{2})\rho_{\text{in}~m+n_\text{loss},n+n_\text{loss}}O(\mu^\frac{n}{2})\nonumber\\
&=O(\mu^{\frac{m+n}{2}}),
\end{align}
where the second line is due to the requirement that $k-m=n_\text{loss}$ and $l-n=n_\text{loss}$ from non-zero Kraus operator elements.


We can now apply the basis transform due to the unitary as described in Eq. \ref{transform}:
\begin{align}
\hat{\rho}_\text{out}&=U\hat{\rho}_\text{lossy}U^\dag\nonumber\\
&=U\left(\sum_{m,n=0}^{n_{\text{max}}}\vert m\rangle\rho_{\text{lossy}~m,n}\langle n\vert \right)U^\dag.
\end{align}
In index notation in the bipartition number state basis,
\begin{align}
    &\rho_{\text{out}~k_\text{u},k_\text{d};k'_\text{u},k'_\text{d}}=\langle k_\text{u},k_\text{d}\vert\hat{\rho}_\text{out}\vert k'_\text{u},k'_\text{d}\rangle\nonumber\\
    =&\sum_{m,n=0}^{n_{\text{max}}}\langle k_\text{u},k_\text{d}\vert U\vert m\rangle\rho_{\text{lossy}~m,n}\langle n\vert U^\dag\vert k'_\text{u},k'_\text{d}\rangle.
\end{align}
Substituting Eq.\ref{kun} and \ref{lossy} into the above expression yields
\begin{align}\label{out}
    &\rho_{\text{out}~k_\text{u},k_\text{d};k'_\text{u},k'_\text{d}}\nonumber\\
    =&\sum_{n_{\text{loss}}=0}^{n_{\text{max}}}\rho_{\text{in}~k_u+k_d+n_\text{loss},k_u'+k_d'+n_\text{loss}}O(\mu^\frac{k_\text{u}+k_\text{d}+k'_\text{u}+k'_\text{d}}{2}).
\end{align}




To compute the contribution to the full system MPO entanglement entropy from $\hat{\rho}_{j,\text{out}}$, we need to vectorize the density operator to obtain $\vert \hat{\rho}_{j,\text{out}}\rangle\rangle$ so that we can pretend it is a pure state and compute its entanglement entropy. The standard procedure of computing the entanglement entropy of a pure state is to obtain the density operator by taking the outer product, obtain the reduced density operator by taking the partial trace over one subsystem, find the reduced density operator's eigenvalues, and take the log average of the eigenvalues. The only difference for the MPO entanglement entropy is that our `pure' state is actually a vectorized density operator, and the eigenvalues may not be normalized since the vectorized state is not $L^2$ normalized.

Vectorization of the density operator, which corresponds to flattening of the matrix and changing bras into kets, is defined as:
\begin{align}
&\hat{\rho}_\text{out}=\sum_{k_\text{u},k_\text{d},k'_\text{u},k'_\text{d}}\vert k_\text{u},k_\text{d}\rangle\rho_{\text{out}~k_\text{u},k_\text{d};k'_\text{u},k'_\text{d}}\langle k'_\text{u},k'_\text{d}\vert\nonumber\\
\rightarrow & \vert\hat{\rho}_\text{out}\rangle\rangle=\sum_{k_\text{u},k_\text{d},k'_\text{u},k'_\text{d}}\rho_{\text{out}~k_\text{u},k_\text{d};k'_\text{u},k'_\text{d}}\vert k_\text{u},k_\text{d};k'_\text{u},k'_\text{d}\rangle\nonumber\\
=&\sum_{K_\text{u},K_\text{d}}\rho_{\text{out}~K_\text{u},K_\text{d}}\vert K_\text{u},K_\text{d}\rangle,
\end{align}
where $K_\text{side}$ is defined as the combined index of $k_\text{side}$ and $k'_\text{side}$. Next, we take the partial trace of its outer product over one bipartition to obtain $\hat{\rho}'=\text{tr}_{\text{u}}(\vert \hat{\rho}_{j,\text{out}}\rangle\rangle\langle\langle\hat{\rho}_{j,\text{out}}\vert)$, yielding
\begin{align}
\rho'_{K_\text{d},\bar{K}_\text{d}}&=\sum_{K_\text{u}}\rho_{\text{out}~K_\text{u},K_\text{d}}~\rho_{\text{out}~K_\text{u},\bar{K}_\text{d}}^*\nonumber\\
&=\sum_{k_\text{u},k'_\text{u}}O(\mu^\frac{k_\text{u}+k_\text{d}+k'_\text{u}+k'_\text{d}}{2})O(\mu^\frac{k_\text{u}+\bar{k}_\text{d}+k'_\text{u}+\bar{k}'_\text{d}}{2})\nonumber\\
&=O(\mu^\frac{k_\text{d}+k'_\text{d}+\bar{k}_\text{d}+\bar{k}'_\text{d}}{2}),
\end{align}
where $\bar{K}$ is the dual of $K$, and $k_\text{u},k'_\text{u}=0$ terms are dominant.

The above analysis is general and independent of the input states. From now on, we will use the fact that the input state is a squeezed state. In GBS, $\rho_{\text{in}~m,n}=0$ if either $m$ or $n$ is odd. Therefore, looking at Eq. \ref{out}, $k_u+k_d+n_\text{loss}$ and $k_u'+k_d'+n_\text{loss}$ (or, more concisely, $k_u+k_d$ and $k_u'+k_d'$) must have the same parity for $\rho_{\text{in}~k_u+k_d+n_\text{loss},k_u'+k_d'+n_\text{loss}}$ to be non-zero. This means that we do not have to consider terms like $\rho_{\text{out}~0,0;0,1},\rho_{\text{out}~1,0;0,0}$, etc. As a result, no half-integer powers of $\mu$ occur in any terms of the output density operator $\hat{\rho}_\text{out}$ or the reduced density operator $\hat{\rho}'$ of the vectorized state.

In this case, it turns out that $\hat{\rho}'$ has exactly one constant order eigenvalue, no first order eigenvalues, and all other eigenvalues are at least second order. To show this, it is sufficient to find all eigenvalues to the first order. Let us write down the form of $\hat{\rho}'$ to the first order, with the first row corresponding to $K_\text{d}=k_\text{d}=k'_\text{d}=0$ and the first column corresponding to $\bar{K}_\text{d}=\bar{k}_\text{d}=\bar{k}'_\text{d}=0$:
\begin{equation}
    \hat{\rho}'=\begin{bmatrix}
    \hat{\rho}_{1,1}' & \hat{\rho}_{1,2}' & \hat{\rho}_{1,3}' & \cdots \\
    \hat{\rho}_{1,2}'^* & 0          & 0          & \vdots \\
    \hat{\rho}_{1,3}'^* & 0          & 0          & \vdots \\
    \vdots     & \cdots     & \cdots     & \ddots 
\end{bmatrix},
\end{equation}
which has eigenvalues
\begin{equation}
    \lambda^2=\frac{1}{2}\left(\hat{\rho}_{1,1}'\pm\sqrt{\hat{\rho}_{1,1}'^2+4\sum_{n=2}^{(n_{\text{max}}+1)^2}\vert \hat{\rho}_{1,n}'\vert ^2}\right),
\end{equation}
and all other eigenvalues are $0$. Note that we call the singular values of the Schmidt decompositions $\lambda$ and the eigenvalues of the reduced density matrices $\lambda^2$. However, $\vert \hat{\rho}_{1,n}'\vert ^2$ is at least $O(\mu^2)$, and the Taylor expansion of the square root will be dominated by the constant and first order contributions from $\hat{\rho}_{1,1}'$. Therefore, the only non-zero first order eigenvalue is $\lambda_1^2=\hat{\rho}_{1,1}'$, which is $O(1)$. 

The above analysis shows that to the second order, the eigenvalues are
\begin{equation}
    \{a+b\mu+c\mu^2,O(\mu^2),O(\mu^2),O(\mu^2),\cdots\}.
\end{equation}
After normalization of the eigenvalues, the entropy contribution due to $\lambda_1^2$ is
\begin{align}
    &-\frac{a+b\mu+c_1\mu^2}{C}\log{\frac{a+b\mu+c_1\mu^2}{C}}\nonumber\\
    &=\frac{(d-c_1)\mu^2}{a\ln{2}}+O(\mu^3)=O(\mu^2),
\end{align}
where $c_i$ is the second order coefficient of $\lambda_i^2$, $d=\sum_{i} c_i$, and $C=a+b\mu+d\mu^2$ is the normalization that must be treated explicitly and not as as a constant. Contribution of other eigenvalues are
\begin{equation}
    -\frac{c_i \mu^2}{C}\log_2{\frac{c_i \mu^2}{C}}=O(-\mu^2\log_2{\mu}).
\end{equation}
Overall, the entanglement entropy scales as $O(\mu^2\log{\mu})$. We would like to understand the scaling of the MPO EE under various loss scalings with the number of input optical modes. Generically, one can consider the situation where the number of surviving photons scales as $N_\text{out}\propto N^\gamma$, making the transmission rate $\mu=\beta N^\gamma/N$. Since the total entanglement entropy is the sum of all $N$ input modes, we obtain
\begin{align}
    S_1(\vert \hat{\rho}\rangle\rangle)&=O\left(N\left(\frac{\beta N^\gamma}{N}\right)^2\log_2{\left(\frac{\beta N^\gamma}{N}\right)}\right)\nonumber\\
    &=O(N^{2\gamma-1}\log_2{N}).
\end{align}

Similarly, for the R\'enyi entropy, contribution from a single $\hat{\rho}_{j,\text{out}}$ is
\begin{align}
&\frac{1}{1-\alpha}\log_2\left[\left(\frac{a+b\mu+c_1\mu^2}{C}\right)^\alpha+\sum_{i\neq 1}\left(\frac{c_i\mu^2}{C}\right)^\alpha\right]\nonumber\\
\approx&\frac{1}{(1-\alpha)\ln{2}}\left(-\frac{d-c_1}{a}\alpha \mu^2+\frac{1}{a}\sum_{i\neq 1}c_i^\alpha \mu^{2\alpha}\right).
\end{align}
Therefore, for $\alpha<1$, the second term dominates, and we have the familiar
\begin{equation}
    S_\alpha=O(N^{1-2(1-\gamma)\alpha}).
\end{equation}
Similarly, for $\alpha>1$, the first term dominates, and we have
\begin{equation}
    S_\alpha=O(\frac{\alpha}{1-\alpha}N^{2\gamma-1}).
\end{equation}

\section{Method of estimating asymptotic MPO EE}\label{app:asymptotic}

For our numerical estimates of GBS operator EE for very large system sizes where direct MPO simulations are impractical, we use the input wavefunction for a single squeezed mode
\begin{align}
    \vert \psi_{\text{in}}\rangle=&\frac{1}{\sqrt{\cosh{r}}}\sum_{n=0}^{n_{\text{max}}/2}\frac{\tanh^n{r}}{2^n n!}\hat{a}^{\dag 2n}\vert 0\rangle,
\end{align}
follow the steps discussed above to compute the EE contribution from a single $\hat{\rho}_{j,\text{out}}$, and multiply by the number of input modes. Specifically, numerically we apply the photon loss Kraus operators defined earlier to this state (both the Kraus operators and the state are truncated to local Hilbert space dimension $n_\text{max}$). We then apply the basis transform due to E.q. \ref{transform}, and the operator is numerically stored in the bipartition $\{\vert k\rangle_\text{u},\vert k \rangle_\text{d}\}$ basis. This transforms an $n_\text{max}\times n_\text{max}$ matrix into a $2n_\text{max}\times 2n_\text{max}$ matrix. This is then followed by vectorization, taking the partial trace of the outer product, fiding the normalized eigenvalues, and computing the entropy.

The coefficients $\cos{\theta_j},\sin{\theta_j}$ related to the unitary can also be simply approximated as $1/\sqrt{2}$ in the asymptotic limit of large $M$ if $l$ is chosen to be $M/2$ as can be seen from E.q. \ref{normalization}. This choice of bipartition is justifiable because we are only interested in the largest EE bipartition since it implies a high simulation cost.

\section{$U(1)$ symmetric tensor network}\label{app:u1}

In an MPS, the probability amplitude tensor $c_{i_1,\dots,i_M}$ can be represented as
\begin{align}
    c_{i_1,\dots,i_M} = \sum_{\alpha_0,\dots,\alpha_M=0}^{\chi-1} &\Gamma_{\alpha_0\alpha_1}^{[1]i_1}\lambda_{\alpha_1}^{[1]}\Gamma_{\alpha_1\alpha_2}^{[2]i_2}\lambda_{\alpha_2}^{[2]}\nonumber\\
    &\dots\lambda_{\alpha_{M-1}}^{[M-1]}\Gamma_{\alpha_{M-1}\alpha_M}^{[M]i_M},
\end{align}
where $\chi$ is a free parameter called the bond dimension, which determines the accuracy of the approximate representation. For systems with higher entanglement, the required bond dimension to achieve a certain fidelity is higher. Overall, the memory complexity of the MPS is determined by the $M$ $\Gamma$ matrices which has $O(\chi^2)$ complexity. This results in an overall memory complexity that is linear in $M$, contrary to the exponential cost of the exact state vector.

Intuitively speaking, the physical index $i_k$ of $\Gamma^{[k]}$ captures the physical degree of free of particle $k$. Further, neighboring $\Gamma^{[k]}$ and $\Gamma^{[k+1]}$ tensors share a dummy index $\alpha_k$ which is contracted, capturing the entanglement between the two neighboring particles. The $\lambda$ tensors correspond to the Schmidt coefficients if a Schmidt decomposition is performed on the wavefunction, and therefore capture entanglement. To apply a local unitary update on particle $k$ and $k+1$, the unitary matrix needs to be contracted with the wavefunction at the physical indices, leading to the resulting tensor
\begin{align}
    \Theta_{\alpha_{k-1}\alpha_{k+1}}^{j_k,j_{k+1}}=
    \sum_{i_k,i_{k+1}=0}^{d-1}\sum_{\alpha_k=0}^{\chi-1}
    &U^{j_k,j_{k+1}}_{i_k,i_{k+1}}\lambda_{\alpha_{k-1}}^{[k-1]}\Gamma_{\alpha_{k-1}\alpha_k}^{[k]i_k}\nonumber\\
    &\lambda_{\alpha_k}^{[k]}\Gamma_{\alpha_k\alpha_{k+1}}^{[k+1]i_{k+1}}\lambda_{\alpha_{k+1}}^{[k+1]},
\end{align}
where the lower and upper indices of $U$ represent input and output degrees of freedom, respectively.

The result of this computation is a single tensor of size $d^2\chi^2$, which should be used in the new representation of the wavefunction to replace $\lambda^{[k-1]},\Gamma^{[k]i_k},\lambda^{[k]},\Gamma^{[k+1]i_{k+1}},\lambda^{[k+1]}$. However, this is no longer in the form of an MPS. To restore the MPS form, singular value decomposition (SVD) is performed on $\Theta$ to produce
\begin{equation}
 \Theta_{\alpha_{k-1}\alpha_{k+1}}^{j_k,j_{k+1}}=\sum_{\beta_k=0}^{d\chi-1}V_{(j_k,\alpha_{k-1}),\beta_k}\Tilde{\lambda}^{[k]}_{\beta_k}W_{\beta_k,(j_{k+1},\alpha_{k+1})}.
\end{equation}
By retaining only the $\chi$ largest singular values, we can identity new $\Gamma$ tensors as
\begin{align}
    \Tilde{\Gamma}_{\alpha_{k-1}\alpha_k}^{[k]i_k}&=V_{(j_k,\alpha_{k-1}),\beta_k}/\lambda_{\alpha_{k-1}}^{[k-1]}\\
    \Tilde{\Gamma}_{\alpha_k\alpha_{k+1}}^{[k+1]i_{k+1}}&=W_{\beta_k,(j_{k+1},\alpha_{k+1})}/\lambda_{\alpha_{k+1}}^{[k+1]},
\end{align}
which restores the MPS form.


Although MPS can efficiently represent many-body systems with controlled entanglement, it does not utilize any symmetry to further reduce the computational cost. To efficiently simulate $U(1)$ symmetric systems, we need to modify the MPS formalism \cite{huang2019simulating,guo2019matrix,oh2021classical}.

We denote the total number of particles to the right of position $k$ corresponding to bond $\alpha_k$ as $c_{\alpha_k}^{[k]}$, then the probability amplitude tensor can be expressed as
\begin{align}
    c_{i_1,\dots,i_M} = \sum_{\alpha_0,\dots,\alpha_M=0}^{\chi-1} &\Gamma_{\alpha_0\alpha_1}^{[1]}\lambda_{\alpha_1}^{[1]}\Gamma_{\alpha_1\alpha_2}^{[2]}\dots
    \lambda_{\alpha_{M-1}}^{[M-1]}\Gamma_{\alpha_{M-1}\alpha_M}^{[M]}\nonumber\\
    \prod_{k=1}^{M}&\delta\left(c_{\alpha_{k-1}}^{[k-1]}-c_{\alpha_k}^{[k]}-i_k\right).
\end{align}
The $\delta$ function essentially determines the correct local particle number based on the charge value difference. Updating the wavefunction according to the unitary can be done with the following procedure. We first realize that a local two-site update does not change the charges at $k-1$ or $k+1$, and we can therefore compute the results for different resulting values of $c^{[k]}$. For each chosen value of $c^{[k]}$, $c^{[k-1]}\geq c^{[k]}$ and $c^{[k+1]}\leq c^{[k]}$, we can select a subset of bonds $\alpha_{k-1}\in\mathcal{A}_{k-1},\alpha_k\in\mathcal{A}_{k},\alpha_{k+1}\in\mathcal{A}_{k+1}$ that satisfy the conditions on the three charges. We can then obtain the $\Theta$ tensor similar to the normal MPS algorithm:
\begin{align}
    &\Theta_{\alpha_{k-1}\alpha_{k+1}}(c^{[k]})=\nonumber\\
    &\sum_{\substack{i_k,i_{k+1}=0\\j_k,j_{k+1}=0}}^{d-1}\sum_{\alpha_k\in\mathcal{A}_k}
    U_{j_k,j_{k+1}}^{i_k,i_{k+1}}\lambda_{\alpha_{k-1}}^{[k-1]}\Gamma_{\alpha_{k-1}\alpha_k}^{[k]}\lambda_{\alpha_k}^{[k]}\Gamma_{\alpha_k\alpha_{k+1}}^{[k+1]}\lambda_{\alpha_{k+1}}^{[k+1]}\nonumber\\
    &\times\delta\left(c_{\alpha_{k-1}}^{[k-1]}-c_{\alpha_k}^{[k]}-j_k\right)\delta\left(c_{\alpha_{k}}^{[k]}-c_{\alpha_{k+1}}^{[k+1]}-j_{k+1}\right)\nonumber\\
    &\times\delta\left(c_{\alpha_{k-1}}^{[k-1]}-c^{[k]}-i_k\right)\delta\left(c^{[k]}-c_{\alpha_{k+1}}^{[k+1]}-i_{k+1}\right)\label{theta},
\end{align}
where $0\leq c^{[k]}\leq N$ and the $\delta$ function determines which entry of the unitary matrix to look up. Additionally, examining the $U(1)$ symmetric MPS tells us that the $\Gamma$ tensors lost their $i$ indices corresponding to the physical degree of freedom (local particle number), reducing the memory complexity by a factor of $d$. This is instead captured by the size $\chi$ 1-d charge tensors $c$. Second, the size of the $\Theta$ matrices that we decompose with SVD is also reduced to at most $\chi\times\chi$ instead of $\chi d\times\chi d$.

We similarly need to compress the new two-site tensor and restore the MPS representation as in the regular algorithm. The full $\Theta$ matrix capturing the result of the contraction should be a block-diagonal matrix with $\Theta(c^{[k]})$'s as composing blocks. Therefore, performing SVD on the full matrix can be achieved by decomposing individual $\Theta(c^{[k]})$, which is the source of the computational complexity reduction of the $U(1)$ symmetric algorithm. Our algorithm performs SVD on all $\Theta(c^{[k]})$'s and keep the $\chi$ largest singular values.

State-of-the-art simulations using tensor networks typically employ hardware acceleration, including the use of novel hardware platforms such as graphical processing units (GPUs) \cite{nguyen2022tensor,lyakh2022exatn,lykov2021performance, lykov2022tensor}. However, symmetry-preserving tensor network algorithms \cite{huang2019simulating,singh2011tensor,guo2019matrix} are highly specialized and require data-dependent array entry look up for the unitary matrix. This is an unusual requirement that is not commonly needed in normal tensor operations such as contraction, reshaping, index permutation, etc. As a result, no highly optimized hardware acceleration is readily available for our algorithm. In this work, we aim to bridge this gap in hardware acceleration for the algorithm by optimizing a subroutine on GPU, and also target our implementation to supercomputing resources.

It is hard to improve SVD as it is a well-researched and optimized routine. The naive implementation of computing $\Theta$ also requires looping over all possible values of $c^{[k-1]}$ and $c^{[k+1]}$, which introduces an additional $O(d^2)$ complexity compared to SVD. Therefore, we focus our discussion on the $\Theta$ computation subroutine and how we optimize it.

\subsection{CPU implementation}

For a given center charge $c^{[k]}$, the CPU-based implementation loops through all possible left and right charge values $c^{[k-1]}$ and $c^{[k+1]}$ and selects a subset of left and right bonds $\alpha_{k-1}$ and $\alpha_{k+1}$ that satisfy the charge requirement. Since $c^{[k]}$ is fixed for each $\Theta(c^{[k]})$ submatrix, the only term that the delta function affects given the charges is $U$ through $j_k, j_{k+1}, i_k, i_{k+1}$. With the correct unitary matrix value identified, the remaining computation is simply tensor contraction. Each iteration partially fills the $\Theta(c^{[k]})$ matrix at bonds $\alpha_{k-1},\alpha_{k+1}$. Iterating over all possible left and right charges fills the entire matrix.

\begin{figure} [ht]
   \begin{center}
   \includegraphics[width=8cm]{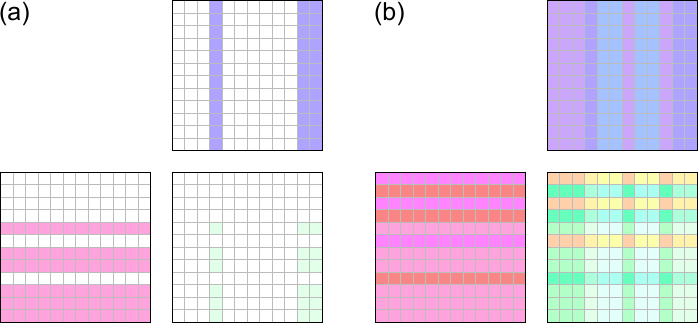}
   \end{center}
   \caption{Algorithms for computing $\Theta$ matrices. (a) CPU-based implementation. A subset of bonds are selected from $\Gamma^{[k]},\Gamma^{[k+1]}$ that have the correct selected charge values $c^{[k-1]}, c^{[k+1]}$. A subset of $\Theta$ is computed. (b) GPU-based implementation. All bonds are used and the entire $\Theta$ matrix is computed at once.}
   { \label{Theta}
}
   \end{figure}

For large total particle number $d$, the $O(d^2)$ complexity due to the nested loop can significantly increase the computational time. Tensor contraction calculations that would otherwise be parallel has to be broken down into pieces. Therefore, the ability to parallelize across different left and right charges and unitary matrix entries is highly desirable, which is exactly what our GPU algorithm accomplishes. The differences between the CPU and GPU implementations are illustrated in Fig. \ref{Theta}.

\subsection{Hierarchical GPU implementation}

A naive parallel implementation of $\Theta$ matrix computation would assign the computation of a single array entry to a single thread. For example, the $\Theta_{i,j}$ can be calculated by a single thread that computes the inner product between the $i$th row of the first matrix and the $j$th column of the second matrix. However, this approach has several limitations, and a non-trivial hierarchical algorithm is used in reality for matrix multiplication. For a pedagogical introduction to the hierarchical approach in the context of matrix multiplication, see the work by Kerr {\sl et. al.} \cite{kerr}

Consider multiplication of $A\in \mathbb{C}^{M\times K}$ and $B\in \mathbb{C}^{K\times N}$. The two matrices are stored in the \textit{global memory} of the GPU, which every thread can access at any time. During inner product calculation of a single thread, the thread needs to read the global memory $2K$ times to complete the row and column vectors. Computing the whole matrix requires $2MNK$ reads of global memory, which turns out to be a limiting factor. Global memory is physically located far away from the compute cores of the GPU, and only a limited amount of memory can be fetched per second. The naive implementation would actually starve the compute cores due to a lack of data, leaving them idling most of the time.

\begin{figure} [ht]
   \begin{center}
   \includegraphics[width=8cm]{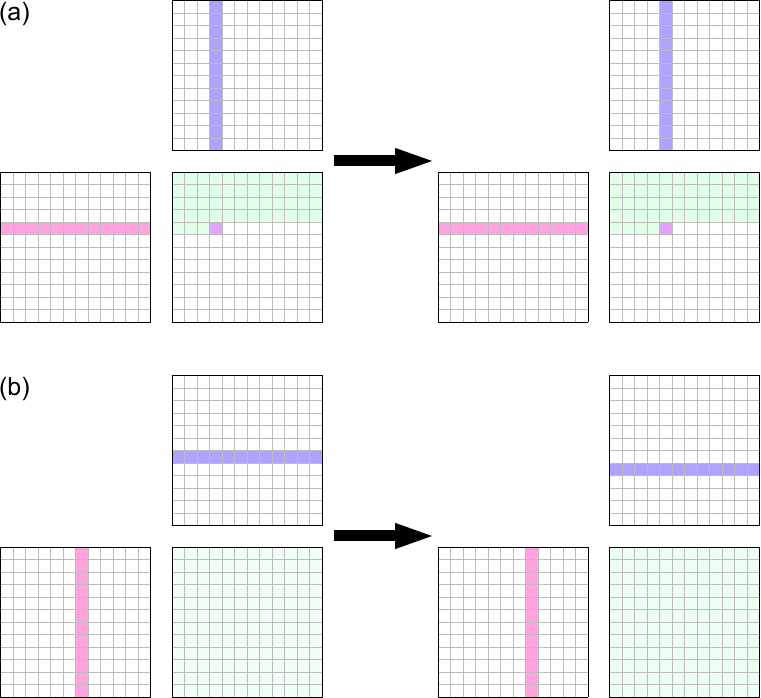}
   \end{center}
   \caption{Matrix multiplication with (a) inner products and (b) outer products.}
   { \label{matmul}
}
   \end{figure}

Alternatively, we can replace element-wise inner products with the accumulation of outer products to reduce the memory read requirement, and Fig. \ref{matmul} illustrates the differences between the two approaches. If a whole row/column of $A/B$ is saved in some memory that is closer to the compute cores but have less capacity, all the threads can accumulate $A_{i,k}B_{k,j}$ once with an outer product. This can be repeated $K$ times to complete matrix multiplication. On a GPU, this closer memory is called \textit{shared memory}, which is shared by threads in its thread block of at most 2048 threads. Each thread block has its owon shared memory. Each outer product requires transfer of data from global to shared memory with $M+N$ global reads, and the entire algorithm only needs $K(M+N)$ global reads and $2MNK$ shared memory reads. In reality, since a thread block has a limited number of threads and shared memory, we cannot fit everything in a single block and must compute the entire output matrix by sub-blocks.

The strategy of shifting the need for high memory access from large capacity broad access slow memory to small capacity local access fast memory can be repeated on lower levels. At the lowest level, a single thread actually computes multiple entries of the matrix, where data is stored in \textit{registers} which are the fastest memory available and are private to each thread. Our GPU algorithm for the $\Theta$ computation subroutine only differs from matrix multiplication by $U$ and $\lambda$ value look up. Therefore, our implementation adopts all the techniques mentioned above to maximize performance.

\subsection{Memory alignment in GPU implementation}

The charge data-dependent access of $U$ poses difficulties in efficient GPU parallelization. In optimized numerical routines, threads access memory in an \textit{aligned} manner, where consecutive threads access consecutive memory addresses, which allows data to be sent in chunks. Sending data chunks allows multiple units of data to be sent in a single clock cycle, otherwise only one unit of data is sent in a given cycle. In a GPU, this can lead to a 32-fold memory bandwidth reduction. If the charge values are completely unpredictable, the memory address of $U$ that needs to be accessed will not be aligned.

\begin{figure} [ht]
   \begin{center}
   \includegraphics[width=8cm]{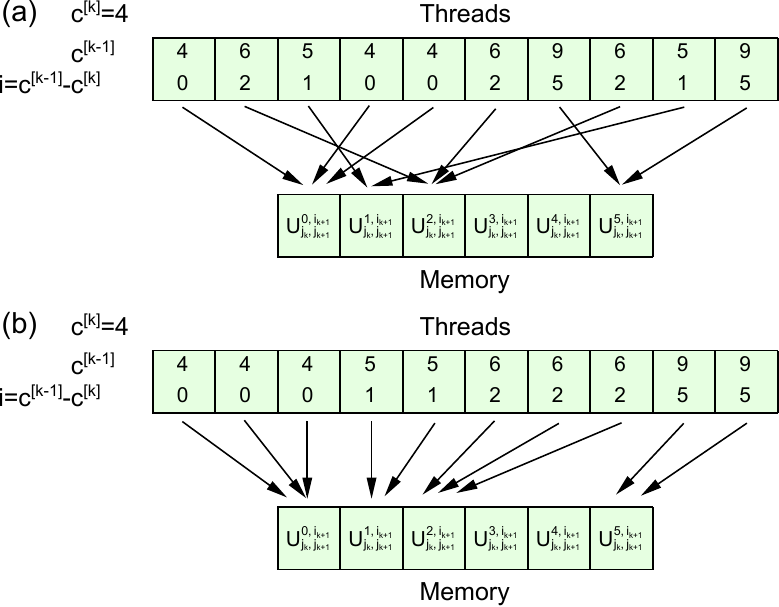}
   \end{center}
   \caption{Memory access pattern (a) without sorting and (b) with sorting.}
   { \label{sort}
}
   \end{figure}

\begin{figure} [ht]
   \begin{center}
   \includegraphics[width=8cm]{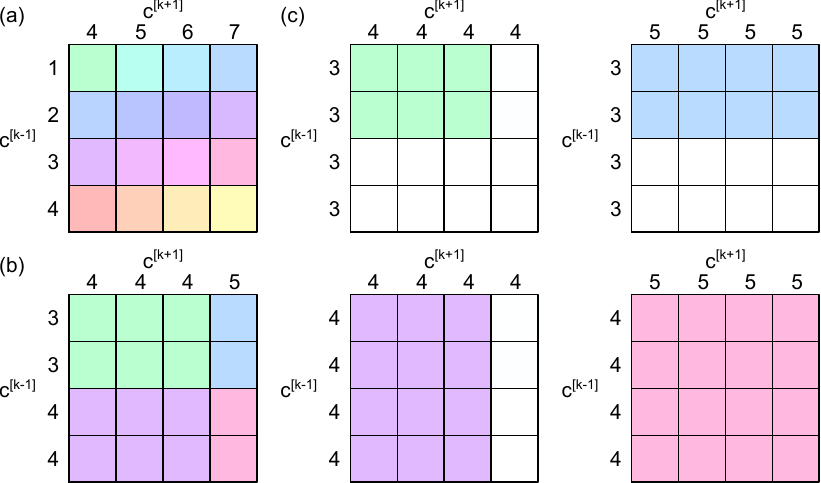}
   \end{center}
   \caption{Illustration of insertion of empty bonds. (a) Worst case scenario of charge value changes within a single fragment without empty bond insertion. The thread has to store 16 values of the unitary. (b) Generic case of a fragment at a charge change boundary. Less than 16 values need to be stored, but this is not known \textit{a-priori} and 16 values of the unitary still needs to be stored. (c) With empty bond insertion, each thread only needs one unitary value.}
   { \label{alignment}
}
   \end{figure}

This issue can be easily addressed by sorting the bonds according to the charge values. This leads to aligned memory access as illustrated in Fig. \ref{sort} and significantly improves performance. Additionally, since each thread calculates multiple entries, it might need to access multiple unitary values even after sorting. Due to the limited number of registers available to each thread, we cannot afford to store redundant unitary values. Therefore, we insert empty bonds to ensure that only one value of the unitary matrix corresponding to a single charge $c$ and physical state $i$ value is stored per thread. This scheme is illustrated in Fig. \ref{alignment}.

Additionally, bond indices are sorted such that $c_{\alpha_k}^{[k]}$ only increases as the bond index increases. For small $d$, this means that $c_{\alpha_k}^{[k]}$ is the same for many consecutive indices. This eliminates the need for threads to look up new $U$ elements, except at boundaries where $c_{\alpha_k}^{[k]}$ changes. This further reduces the need for memory access and reduces latency.

\subsection{High-level parallelization}
Besides the numerical parallelization of individual SVD and $\Theta$ matrix computations through the use of GPUs, additional parallelization is explicitly implemented on the algorithmic level. Further, for systems with large bond dimensions, storing the entire tensor network on a single-GPU or even a single node may become prohibitive. We distribute the storage of individual $\Gamma$ tensors to different nodes.

First, we parallelize independent two-site unitary updates. A host node identifies all parallel local unitary updates and keeps track of a list of available and busy nodes. Local unitary updates are allocated as soon as a node is available. During allocation, the compute process of the computational node requests the needed $\Gamma,\lambda,c$ tensors from the storage processes of the corresponding storage nodes. Similar communication takes place after the computation to update the stored tensors. Second, for a single beam splitter MPO update, the overall $\Theta$ matrix is broken up into $\Theta(c^{[k]})$'s, which we compute and decompose in parallel. After the computation node receives the data needed, the data needed for each $\Theta(c^{[k]})$ is distributed to individual GPUs.

\begin{figure} [ht]
   \begin{center}
   \includegraphics[width=8cm]{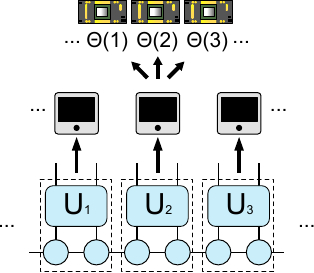}
   \end{center}
   \caption{High-level parallelization. Independent unitary gate updates are distributed to different nodes. Within each unitary update, $\Theta(c^{[k]})$'s are computed and decomposed with SVD independently on different GPUs.}
   { \label{parallel}
}
   \end{figure}

With the high-level parallelization discussed above and illustrated in Fig. \ref{parallel}, the algorithm can be easily scaled to supercomputers with multiple nodes and GPUs, especially when the system size is large. However, there are smaller systems that do not require multi-node parallelization, and we provide implementations with intermediary parallelism as well to avoid the communication overhead of the fully parallel algorithm. On the lowest level, only one GPU is considered, and no distributed memory or computation is used. On the second level, all memory is managed by a single node, and unitary updates are distributed to individual GPUs instead of nodes.

\subsection{Run time reduction}\label{Results}

We evaluate the performance of our GPU supercomputing algorithm against the CPU-only implementation at the Argonne Leadership Computing Facility
(ALCF). All CPU simulations are performed with a single node of the Bebop system with a 2.10 GHz Intel Xeon E5-2695v4 32-core CPU, and GPU simulations are performed on the Polaris system. A single node of the Polaris system has 4 Nvidia A100 GPUs. Table \textrm{I} shows the simulation time in seconds of different implementations for a lossy boson sampling experiment with 12 modes, 10 input squeezed modes, bond dimension 1024 and 8192, photon loss rate 0.55, and local Hilbert space dimension 15. Increasing the bond dimension $\chi$ increases the simulation accuracy and time. Moreover, lossy boson sampling requires the density matrix instead of the state vector, and the generalized algorithm is described in detail in by Oh {\sl et. al.} \cite{oh2021classical} The consequence of the density matrix generalization is that each charge can take on $15^2=225$ values instead of only $15$, which means that the CPU-based algorithm needs to perform $225^2=50625$ iterations to fill the $\Theta$ matrix. On the other hand, our GPU algorithm computes all entries of $\Theta$ in parallel.

For the small $\chi$ experiment, we ar able to simulate using only CPUs in a reasonable amount of time for comparison. Encouragingly, the single-GPU algorithm achieves a dramatic 63-time speedup even with a single-GPU. We further test the unitary-level parallel algorithm on one node and observe a further two-fold speedup. Lastly, we use the fully parallel algorithm on 6 nodes, observing an additional 43\% increase in the computational speed. We observe that the gain in computational speed by switching from less parallelized to highly parallelized implementation is less than the increase in computational resources. Higher-level parallelism incurs significant overhead, which indicates that there is still significant room for optimization.

Fortunately, this payoff in higher parallelism is more pronounced in the setting of larger system sizes. The CPU implementation failed to complete the simulation within the maximum allowed wall time of $72$ hours. This means that our single-GPU implementation achieves at least a 125-fold speed up. The computational time is further reduced two-fold when going from the single-GPU implementation to the unitary-level parallel algorithm on one node, similar to the small bond dimension case. However, changing to the fully parallelized algorithm with 6 nodes further reduces the time more than three-fold compared to a fractional reduction in the small bond dimension case. Overall, the fully parallel implementation on six nodes is on the order of a 1000 times faster than the 32-core CPU implementation.

\begin{table}
\centering
\begin{tabular}{ |c|c|c|c|c| }
 \hline
 & CPU & single-GPU & One node & Six nodes \\ 
 \hline
$\chi=1024$ & 7966 & 126 & 60 & 42 \\
 \hline
 $\chi=8192$ & >259000 & 2066 & 1045 & 322 \\
 \hline
\end{tabular}
\caption{Simulation time in seconds.}
\end{table}

\begin{figure} [ht]
   \begin{center}
   \includegraphics[width=8cm]{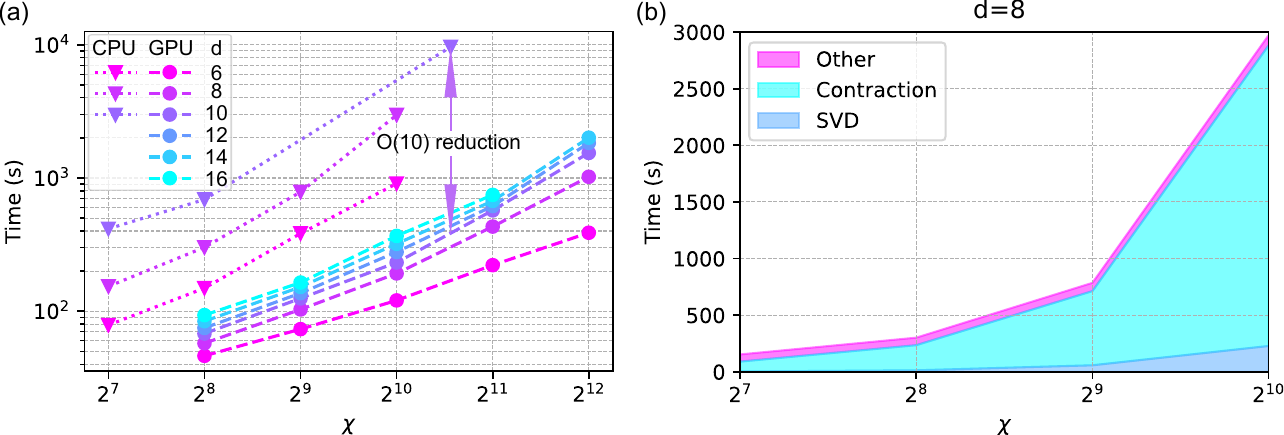}
   \end{center}
   \caption{(a) CPU and GPU simulation time. (b) Contribution to the CPU simulation time from subroutines.}
   { \label{GPU_vs_CPU_timing}
}
   \end{figure}

\begin{figure} [ht]
   \begin{center}
   \includegraphics[width=8cm]{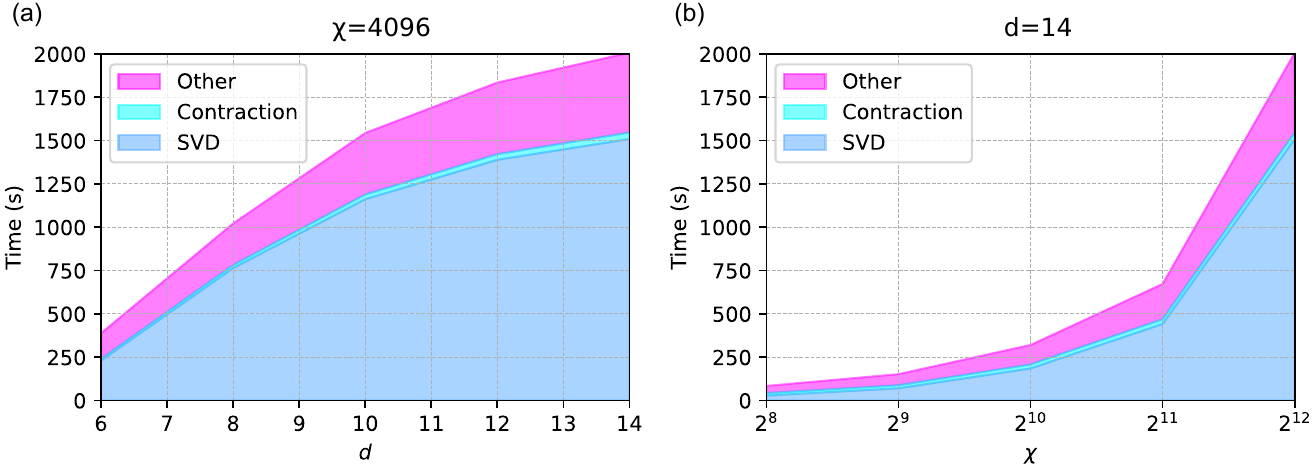}
   \end{center}
   \caption{Contribution to simulation time of the GPU algorithm from subroutines. (a) Bond dimension $\chi=4096$. (b) Local Hilbert space size $d=14$.}
   { \label{GPU_time_contribution}
}
   \end{figure}

The exact speed up depends on the system size, so we show more experiments with different configurations. All the following experiments are performed with $N=5,M=32,\mu=0.5,r=0.88$ on a single 32-core CPU or a single A100 GPU. We show the CPU and GPU simulation time for various systems in Fig. \ref{GPU_vs_CPU_timing}a. For system sizes that the CPU can reasonably complete, we observe over 10 times speed up on a single GPU. Further, Fig. \ref{GPU_vs_CPU_timing}b shows that tensor contraction time dominates the overall run time due to the highly inefficient double nested loop. Fig. \ref{GPU_time_contribution} shows contributions to the GPU simulation time from subroutines.  We see that the contribution from the tensor contraction step is minimal compared to the total simulation time thanks to the efficient custom kernel. Further, SVD takes up the majority of the simulation time, meaning that the overhead of tensor sorting, alignment, and storage is acceptable.

\bibliography{boson.bib}

\begin{thebibliography}{74}%
\makeatletter
\providecommand \@ifxundefined [1]{%
 \@ifx{#1\undefined}
}%
\providecommand \@ifnum [1]{%
 \ifnum #1\expandafter \@firstoftwo
 \else \expandafter \@secondoftwo
 \fi
}%
\providecommand \@ifx [1]{%
 \ifx #1\expandafter \@firstoftwo
 \else \expandafter \@secondoftwo
 \fi
}%
\providecommand \natexlab [1]{#1}%
\providecommand \enquote  [1]{``#1''}%
\providecommand \bibnamefont  [1]{#1}%
\providecommand \bibfnamefont [1]{#1}%
\providecommand \citenamefont [1]{#1}%
\providecommand \href@noop [0]{\@secondoftwo}%
\providecommand \href [0]{\begingroup \@sanitize@url \@href}%
\providecommand \@href[1]{\@@startlink{#1}\@@href}%
\providecommand \@@href[1]{\endgroup#1\@@endlink}%
\providecommand \@sanitize@url [0]{\catcode `\\12\catcode `\$12\catcode
  `\&12\catcode `\#12\catcode `\^12\catcode `\_12\catcode `\%12\relax}%
\providecommand \@@startlink[1]{}%
\providecommand \@@endlink[0]{}%
\providecommand \url  [0]{\begingroup\@sanitize@url \@url }%
\providecommand \@url [1]{\endgroup\@href {#1}{\urlprefix }}%
\providecommand \urlprefix  [0]{URL }%
\providecommand \Eprint [0]{\href }%
\providecommand \doibase [0]{https://doi.org/}%
\providecommand \selectlanguage [0]{\@gobble}%
\providecommand \bibinfo  [0]{\@secondoftwo}%
\providecommand \bibfield  [0]{\@secondoftwo}%
\providecommand \translation [1]{[#1]}%
\providecommand \BibitemOpen [0]{}%
\providecommand \bibitemStop [0]{}%
\providecommand \bibitemNoStop [0]{.\EOS\space}%
\providecommand \EOS [0]{\spacefactor3000\relax}%
\providecommand \BibitemShut  [1]{\csname bibitem#1\endcsname}%
\let\auto@bib@innerbib\@empty
\bibitem [{\citenamefont {Shor}(1994)}]{shor1994algorithms}%
  \BibitemOpen
  \bibfield  {author} {\bibinfo {author} {\bibfnamefont {P.~W.}\ \bibnamefont
  {Shor}},\ }\bibfield  {title} {\bibinfo {title} {Algorithms for quantum
  computation: discrete logarithms and factoring},\ }in\ \href
  {https://doi.org/10.1109/SFCS.1994.365700} {\emph {\bibinfo {booktitle}
  {Proceedings 35th annual symposium on foundations of computer science}}}\
  (\bibinfo {organization} {Ieee},\ \bibinfo {year} {1994})\ pp.\ \bibinfo
  {pages} {124--134}\BibitemShut {NoStop}%
\bibitem [{\citenamefont {Grover}(1996)}]{grover1996a}%
  \BibitemOpen
  \bibfield  {author} {\bibinfo {author} {\bibfnamefont {L.~K.}\ \bibnamefont
  {Grover}},\ }\bibfield  {title} {\bibinfo {title} {A fast quantum mechanical
  algorithm for database search},\ }in\ \href
  {https://doi.org/10.1145/237814.237866} {\emph {\bibinfo {booktitle}
  {Proceedings of the Twenty-Eighth Annual ACM Symposium on Theory of
  Computing}}},\ \bibinfo {series and number} {STOC '96}\ (\bibinfo
  {publisher} {Association for Computing Machinery},\ \bibinfo {address} {New
  York, NY, USA},\ \bibinfo {year} {1996})\ p.\ \bibinfo {pages}
  {212–219}\BibitemShut {NoStop}%
\bibitem [{\citenamefont {Harrow}\ \emph {et~al.}(2009)\citenamefont {Harrow},
  \citenamefont {Hassidim},\ and\ \citenamefont {Lloyd}}]{harrow2009quantum}%
  \BibitemOpen
  \bibfield  {author} {\bibinfo {author} {\bibfnamefont {A.~W.}\ \bibnamefont
  {Harrow}}, \bibinfo {author} {\bibfnamefont {A.}~\bibnamefont {Hassidim}},\
  and\ \bibinfo {author} {\bibfnamefont {S.}~\bibnamefont {Lloyd}},\ }\bibfield
   {title} {\bibinfo {title} {Quantum algorithm for linear systems of
  equations},\ }\href {https://doi.org/10.1103/PhysRevLett.103.150502}
  {\bibfield  {journal} {\bibinfo  {journal} {Phys. Rev. Lett.}\ }\textbf
  {\bibinfo {volume} {103}},\ \bibinfo {pages} {150502} (\bibinfo {year}
  {2009})}\BibitemShut {NoStop}%
\bibitem [{\citenamefont {Berry}\ \emph {et~al.}(2007)\citenamefont {Berry},
  \citenamefont {Ahokas}, \citenamefont {Cleve},\ and\ \citenamefont
  {Sanders}}]{berry2007Efficient}%
  \BibitemOpen
  \bibfield  {author} {\bibinfo {author} {\bibfnamefont {D.~W.}\ \bibnamefont
  {Berry}}, \bibinfo {author} {\bibfnamefont {G.}~\bibnamefont {Ahokas}},
  \bibinfo {author} {\bibfnamefont {R.}~\bibnamefont {Cleve}},\ and\ \bibinfo
  {author} {\bibfnamefont {B.~C.}\ \bibnamefont {Sanders}},\ }\bibfield
  {title} {\bibinfo {title} {Efficient quantum algorithms for simulating sparse
  hamiltonians},\ }\href {https://doi.org/10.1007/s00220-006-0150-x} {\bibfield
   {journal} {\bibinfo  {journal} {Commun. Math. Phys.}\ }\textbf {\bibinfo
  {volume} {270}},\ \bibinfo {pages} {359–371} (\bibinfo {year}
  {2007})}\BibitemShut {NoStop}%
\bibitem [{\citenamefont {Berry}\ \emph {et~al.}(2014)\citenamefont {Berry},
  \citenamefont {Childs}, \citenamefont {Cleve}, \citenamefont {Kothari},\ and\
  \citenamefont {Somma}}]{berry2014exponential}%
  \BibitemOpen
  \bibfield  {author} {\bibinfo {author} {\bibfnamefont {D.~W.}\ \bibnamefont
  {Berry}}, \bibinfo {author} {\bibfnamefont {A.~M.}\ \bibnamefont {Childs}},
  \bibinfo {author} {\bibfnamefont {R.}~\bibnamefont {Cleve}}, \bibinfo
  {author} {\bibfnamefont {R.}~\bibnamefont {Kothari}},\ and\ \bibinfo {author}
  {\bibfnamefont {R.~D.}\ \bibnamefont {Somma}},\ }\bibfield  {title} {\bibinfo
  {title} {Exponential improvement in precision for simulating sparse
  hamiltonians},\ }in\ \href {https://doi.org/10.1145/2591796.2591854} {\emph
  {\bibinfo {booktitle} {Proceedings of the Forty-Sixth Annual ACM Symposium on
  Theory of Computing}}},\ \bibinfo {series and number} {STOC '14}\ (\bibinfo
  {publisher} {Association for Computing Machinery},\ \bibinfo {address} {New
  York, NY, USA},\ \bibinfo {year} {2014})\ p.\ \bibinfo {pages}
  {283–292}\BibitemShut {NoStop}%
\bibitem [{\citenamefont {Childs}(2010)}]{childs2010on}%
  \BibitemOpen
  \bibfield  {author} {\bibinfo {author} {\bibfnamefont {A.}~\bibnamefont
  {Childs}},\ }\bibfield  {title} {\bibinfo {title} {On the relationship
  between continuous- and discrete-time quantum walk},\ }\href
  {https://doi.org/10.1007/s00220-009-0930-1} {\bibfield  {journal} {\bibinfo
  {journal} {Commun. Math. Phys.}\ }\textbf {\bibinfo {volume} {294}},\
  \bibinfo {pages} {581–603} (\bibinfo {year} {2010})}\BibitemShut {NoStop}%
\bibitem [{\citenamefont {Low}\ and\ \citenamefont
  {Chuang}(2017)}]{low2017optimal}%
  \BibitemOpen
  \bibfield  {author} {\bibinfo {author} {\bibfnamefont {G.~H.}\ \bibnamefont
  {Low}}\ and\ \bibinfo {author} {\bibfnamefont {I.~L.}\ \bibnamefont
  {Chuang}},\ }\bibfield  {title} {\bibinfo {title} {Optimal hamiltonian
  simulation by quantum signal processing},\ }\href
  {https://doi.org/10.1103/PhysRevLett.118.010501} {\bibfield  {journal}
  {\bibinfo  {journal} {Phys. Rev. Lett.}\ }\textbf {\bibinfo {volume} {118}},\
  \bibinfo {pages} {010501} (\bibinfo {year} {2017})}\BibitemShut {NoStop}%
\bibitem [{\citenamefont {Alexeev}\ \emph {et~al.}(2021)\citenamefont {Alexeev}
  \emph {et~al.}}]{Alexeev2021}%
  \BibitemOpen
  \bibfield  {author} {\bibinfo {author} {\bibfnamefont {Y.}~\bibnamefont
  {Alexeev}} \emph {et~al.},\ }\bibfield  {title} {\bibinfo {title} {Quantum
  computer systems for scientific discovery},\ }\bibfield  {journal} {\bibinfo
  {journal} {{PRX} Quantum}\ }\textbf {\bibinfo {volume} {2}},\ \href
  {https://doi.org/10.1103/prxquantum.2.017001} {10.1103/prxquantum.2.017001}
  (\bibinfo {year} {2021})\BibitemShut {NoStop}%
\bibitem [{\citenamefont {Aaronson}\ and\ \citenamefont
  {Arkhipov}(2011)}]{aaronson2011computational}%
  \BibitemOpen
  \bibfield  {author} {\bibinfo {author} {\bibfnamefont {S.}~\bibnamefont
  {Aaronson}}\ and\ \bibinfo {author} {\bibfnamefont {A.}~\bibnamefont
  {Arkhipov}},\ }\bibfield  {title} {\bibinfo {title} {The computational
  complexity of linear optics},\ }in\ \href
  {https://doi.org/10.1145/1993636.1993682} {\emph {\bibinfo {booktitle}
  {Proceedings of the forty-third annual ACM symposium on Theory of
  computing}}}\ (\bibinfo {year} {2011})\ pp.\ \bibinfo {pages}
  {333--342}\BibitemShut {NoStop}%
\bibitem [{\citenamefont {Broome}\ \emph {et~al.}(2013)\citenamefont {Broome},
  \citenamefont {Fedrizzi}, \citenamefont {Rahimi-Keshari}, \citenamefont
  {Dove}, \citenamefont {Aaronson}, \citenamefont {Ralph},\ and\ \citenamefont
  {White}}]{broome2013photonic}%
  \BibitemOpen
  \bibfield  {author} {\bibinfo {author} {\bibfnamefont {M.~A.}\ \bibnamefont
  {Broome}}, \bibinfo {author} {\bibfnamefont {A.}~\bibnamefont {Fedrizzi}},
  \bibinfo {author} {\bibfnamefont {S.}~\bibnamefont {Rahimi-Keshari}},
  \bibinfo {author} {\bibfnamefont {J.}~\bibnamefont {Dove}}, \bibinfo {author}
  {\bibfnamefont {S.}~\bibnamefont {Aaronson}}, \bibinfo {author}
  {\bibfnamefont {T.~C.}\ \bibnamefont {Ralph}},\ and\ \bibinfo {author}
  {\bibfnamefont {A.~G.}\ \bibnamefont {White}},\ }\bibfield  {title} {\bibinfo
  {title} {Photonic boson sampling in a tunable circuit},\ }\href
  {https://doi.org/10.1126/science.1231440} {\bibfield  {journal} {\bibinfo
  {journal} {Science}\ }\textbf {\bibinfo {volume} {339}},\ \bibinfo {pages}
  {794} (\bibinfo {year} {2013})}\BibitemShut {NoStop}%
\bibitem [{\citenamefont {Spring}\ \emph {et~al.}(2013)\citenamefont {Spring},
  \citenamefont {Metcalf}, \citenamefont {Humphreys}, \citenamefont
  {Kolthammer}, \citenamefont {Jin}, \citenamefont {Barbieri}, \citenamefont
  {Datta}, \citenamefont {Thomas-Peter}, \citenamefont {Langford},
  \citenamefont {Kundys} \emph {et~al.}}]{spring2013boson}%
  \BibitemOpen
  \bibfield  {author} {\bibinfo {author} {\bibfnamefont {J.~B.}\ \bibnamefont
  {Spring}}, \bibinfo {author} {\bibfnamefont {B.~J.}\ \bibnamefont {Metcalf}},
  \bibinfo {author} {\bibfnamefont {P.~C.}\ \bibnamefont {Humphreys}}, \bibinfo
  {author} {\bibfnamefont {W.~S.}\ \bibnamefont {Kolthammer}}, \bibinfo
  {author} {\bibfnamefont {X.-M.}\ \bibnamefont {Jin}}, \bibinfo {author}
  {\bibfnamefont {M.}~\bibnamefont {Barbieri}}, \bibinfo {author}
  {\bibfnamefont {A.}~\bibnamefont {Datta}}, \bibinfo {author} {\bibfnamefont
  {N.}~\bibnamefont {Thomas-Peter}}, \bibinfo {author} {\bibfnamefont {N.~K.}\
  \bibnamefont {Langford}}, \bibinfo {author} {\bibfnamefont {D.}~\bibnamefont
  {Kundys}}, \emph {et~al.},\ }\bibfield  {title} {\bibinfo {title} {Boson
  sampling on a photonic chip},\ }\href
  {https://doi.org/10.1126/science.1231692} {\bibfield  {journal} {\bibinfo
  {journal} {Science}\ }\textbf {\bibinfo {volume} {339}},\ \bibinfo {pages}
  {798} (\bibinfo {year} {2013})}\BibitemShut {NoStop}%
\bibitem [{\citenamefont {Tillmann}\ \emph {et~al.}(2013)\citenamefont
  {Tillmann}, \citenamefont {Daki{\'c}}, \citenamefont {Heilmann},
  \citenamefont {Nolte}, \citenamefont {Szameit},\ and\ \citenamefont
  {Walther}}]{tillmann2013experimental}%
  \BibitemOpen
  \bibfield  {author} {\bibinfo {author} {\bibfnamefont {M.}~\bibnamefont
  {Tillmann}}, \bibinfo {author} {\bibfnamefont {B.}~\bibnamefont {Daki{\'c}}},
  \bibinfo {author} {\bibfnamefont {R.}~\bibnamefont {Heilmann}}, \bibinfo
  {author} {\bibfnamefont {S.}~\bibnamefont {Nolte}}, \bibinfo {author}
  {\bibfnamefont {A.}~\bibnamefont {Szameit}},\ and\ \bibinfo {author}
  {\bibfnamefont {P.}~\bibnamefont {Walther}},\ }\bibfield  {title} {\bibinfo
  {title} {Experimental boson sampling},\ }\href
  {https://doi.org/https://doi.org/10.1038/nphoton.2013.102} {\bibfield
  {journal} {\bibinfo  {journal} {Nat. Photonics}\ }\textbf {\bibinfo {volume}
  {7}},\ \bibinfo {pages} {540} (\bibinfo {year} {2013})}\BibitemShut {NoStop}%
\bibitem [{\citenamefont {Crespi}\ \emph {et~al.}(2013)\citenamefont {Crespi},
  \citenamefont {Osellame}, \citenamefont {Ramponi}, \citenamefont {Brod},
  \citenamefont {Galvao}, \citenamefont {Spagnolo}, \citenamefont {Vitelli},
  \citenamefont {Maiorino}, \citenamefont {Mataloni},\ and\ \citenamefont
  {Sciarrino}}]{crespi2013integrated}%
  \BibitemOpen
  \bibfield  {author} {\bibinfo {author} {\bibfnamefont {A.}~\bibnamefont
  {Crespi}}, \bibinfo {author} {\bibfnamefont {R.}~\bibnamefont {Osellame}},
  \bibinfo {author} {\bibfnamefont {R.}~\bibnamefont {Ramponi}}, \bibinfo
  {author} {\bibfnamefont {D.~J.}\ \bibnamefont {Brod}}, \bibinfo {author}
  {\bibfnamefont {E.~F.}\ \bibnamefont {Galvao}}, \bibinfo {author}
  {\bibfnamefont {N.}~\bibnamefont {Spagnolo}}, \bibinfo {author}
  {\bibfnamefont {C.}~\bibnamefont {Vitelli}}, \bibinfo {author} {\bibfnamefont
  {E.}~\bibnamefont {Maiorino}}, \bibinfo {author} {\bibfnamefont
  {P.}~\bibnamefont {Mataloni}},\ and\ \bibinfo {author} {\bibfnamefont
  {F.}~\bibnamefont {Sciarrino}},\ }\bibfield  {title} {\bibinfo {title}
  {Integrated multimode interferometers with arbitrary designs for photonic
  boson sampling},\ }\href {https://doi.org/10.1038/nphoton.2013.112}
  {\bibfield  {journal} {\bibinfo  {journal} {Nat. Photonics}\ }\textbf
  {\bibinfo {volume} {7}},\ \bibinfo {pages} {545} (\bibinfo {year}
  {2013})}\BibitemShut {NoStop}%
\bibitem [{\citenamefont {Spagnolo}\ \emph {et~al.}(2014)\citenamefont
  {Spagnolo}, \citenamefont {Vitelli}, \citenamefont {Bentivegna},
  \citenamefont {Brod}, \citenamefont {Crespi}, \citenamefont {Flamini},
  \citenamefont {Giacomini}, \citenamefont {Milani}, \citenamefont {Ramponi},
  \citenamefont {Mataloni} \emph {et~al.}}]{spagnolo2014experimental}%
  \BibitemOpen
  \bibfield  {author} {\bibinfo {author} {\bibfnamefont {N.}~\bibnamefont
  {Spagnolo}}, \bibinfo {author} {\bibfnamefont {C.}~\bibnamefont {Vitelli}},
  \bibinfo {author} {\bibfnamefont {M.}~\bibnamefont {Bentivegna}}, \bibinfo
  {author} {\bibfnamefont {D.~J.}\ \bibnamefont {Brod}}, \bibinfo {author}
  {\bibfnamefont {A.}~\bibnamefont {Crespi}}, \bibinfo {author} {\bibfnamefont
  {F.}~\bibnamefont {Flamini}}, \bibinfo {author} {\bibfnamefont
  {S.}~\bibnamefont {Giacomini}}, \bibinfo {author} {\bibfnamefont
  {G.}~\bibnamefont {Milani}}, \bibinfo {author} {\bibfnamefont
  {R.}~\bibnamefont {Ramponi}}, \bibinfo {author} {\bibfnamefont
  {P.}~\bibnamefont {Mataloni}}, \emph {et~al.},\ }\bibfield  {title} {\bibinfo
  {title} {Experimental validation of photonic boson sampling},\ }\href
  {https://doi.org/10.1038/nphoton.2014.135} {\bibfield  {journal} {\bibinfo
  {journal} {Nat. Photonics}\ }\textbf {\bibinfo {volume} {8}},\ \bibinfo
  {pages} {615} (\bibinfo {year} {2014})}\BibitemShut {NoStop}%
\bibitem [{\citenamefont {Carolan}\ \emph {et~al.}(2014)\citenamefont
  {Carolan}, \citenamefont {Meinecke}, \citenamefont {Shadbolt}, \citenamefont
  {Russell}, \citenamefont {Ismail}, \citenamefont {W{\"o}rhoff}, \citenamefont
  {Rudolph}, \citenamefont {Thompson}, \citenamefont {O'brien}, \citenamefont
  {Matthews} \emph {et~al.}}]{carolan2014experimental}%
  \BibitemOpen
  \bibfield  {author} {\bibinfo {author} {\bibfnamefont {J.}~\bibnamefont
  {Carolan}}, \bibinfo {author} {\bibfnamefont {J.~D.~A.}\ \bibnamefont
  {Meinecke}}, \bibinfo {author} {\bibfnamefont {P.~J.}\ \bibnamefont
  {Shadbolt}}, \bibinfo {author} {\bibfnamefont {N.~J.}\ \bibnamefont
  {Russell}}, \bibinfo {author} {\bibfnamefont {N.}~\bibnamefont {Ismail}},
  \bibinfo {author} {\bibfnamefont {K.}~\bibnamefont {W{\"o}rhoff}}, \bibinfo
  {author} {\bibfnamefont {T.}~\bibnamefont {Rudolph}}, \bibinfo {author}
  {\bibfnamefont {M.~G.}\ \bibnamefont {Thompson}}, \bibinfo {author}
  {\bibfnamefont {J.~L.}\ \bibnamefont {O'brien}}, \bibinfo {author}
  {\bibfnamefont {J.~C.~F.}\ \bibnamefont {Matthews}}, \emph {et~al.},\
  }\bibfield  {title} {\bibinfo {title} {On the experimental verification of
  quantum complexity in linear optics},\ }\href
  {https://doi.org/10.1038/nphoton.2014.152} {\bibfield  {journal} {\bibinfo
  {journal} {Nat. Photonics}\ }\textbf {\bibinfo {volume} {8}},\ \bibinfo
  {pages} {621} (\bibinfo {year} {2014})}\BibitemShut {NoStop}%
\bibitem [{\citenamefont {Carolan}\ \emph {et~al.}(2015)\citenamefont
  {Carolan}, \citenamefont {Harrold}, \citenamefont {Sparrow}, \citenamefont
  {Mart{\'\i}n-L{\'o}pez}, \citenamefont {Russell}, \citenamefont
  {Silverstone}, \citenamefont {Shadbolt}, \citenamefont {Matsuda},
  \citenamefont {Oguma}, \citenamefont {Itoh} \emph
  {et~al.}}]{carolan2015universal}%
  \BibitemOpen
  \bibfield  {author} {\bibinfo {author} {\bibfnamefont {J.}~\bibnamefont
  {Carolan}}, \bibinfo {author} {\bibfnamefont {C.}~\bibnamefont {Harrold}},
  \bibinfo {author} {\bibfnamefont {C.}~\bibnamefont {Sparrow}}, \bibinfo
  {author} {\bibfnamefont {E.}~\bibnamefont {Mart{\'\i}n-L{\'o}pez}}, \bibinfo
  {author} {\bibfnamefont {N.~J.}\ \bibnamefont {Russell}}, \bibinfo {author}
  {\bibfnamefont {J.~W.}\ \bibnamefont {Silverstone}}, \bibinfo {author}
  {\bibfnamefont {P.~J.}\ \bibnamefont {Shadbolt}}, \bibinfo {author}
  {\bibfnamefont {N.}~\bibnamefont {Matsuda}}, \bibinfo {author} {\bibfnamefont
  {M.}~\bibnamefont {Oguma}}, \bibinfo {author} {\bibfnamefont
  {M.}~\bibnamefont {Itoh}}, \emph {et~al.},\ }\bibfield  {title} {\bibinfo
  {title} {Universal linear optics},\ }\href
  {https://doi.org/10.1126/science.aab3642} {\bibfield  {journal} {\bibinfo
  {journal} {Science}\ }\textbf {\bibinfo {volume} {349}},\ \bibinfo {pages}
  {711} (\bibinfo {year} {2015})}\BibitemShut {NoStop}%
\bibitem [{\citenamefont {Bentivegna}\ \emph {et~al.}(2015)\citenamefont
  {Bentivegna}, \citenamefont {Spagnolo}, \citenamefont {Vitelli},
  \citenamefont {Flamini}, \citenamefont {Viggianiello}, \citenamefont
  {Latmiral}, \citenamefont {Mataloni}, \citenamefont {Brod}, \citenamefont
  {Galv{\~a}o}, \citenamefont {Crespi} \emph
  {et~al.}}]{bentivegna2015experimental}%
  \BibitemOpen
  \bibfield  {author} {\bibinfo {author} {\bibfnamefont {M.}~\bibnamefont
  {Bentivegna}}, \bibinfo {author} {\bibfnamefont {N.}~\bibnamefont
  {Spagnolo}}, \bibinfo {author} {\bibfnamefont {C.}~\bibnamefont {Vitelli}},
  \bibinfo {author} {\bibfnamefont {F.}~\bibnamefont {Flamini}}, \bibinfo
  {author} {\bibfnamefont {N.}~\bibnamefont {Viggianiello}}, \bibinfo {author}
  {\bibfnamefont {L.}~\bibnamefont {Latmiral}}, \bibinfo {author}
  {\bibfnamefont {P.}~\bibnamefont {Mataloni}}, \bibinfo {author}
  {\bibfnamefont {D.~J.}\ \bibnamefont {Brod}}, \bibinfo {author}
  {\bibfnamefont {E.~F.}\ \bibnamefont {Galv{\~a}o}}, \bibinfo {author}
  {\bibfnamefont {A.}~\bibnamefont {Crespi}}, \emph {et~al.},\ }\bibfield
  {title} {\bibinfo {title} {Experimental scattershot boson sampling},\ }\href
  {https://doi.org/10.1126/sciadv.1400255} {\bibfield  {journal} {\bibinfo
  {journal} {Sci. Adv.}\ }\textbf {\bibinfo {volume} {1}},\ \bibinfo {pages}
  {e1400255} (\bibinfo {year} {2015})}\BibitemShut {NoStop}%
\bibitem [{\citenamefont {Zhong}\ \emph {et~al.}(2018)\citenamefont {Zhong}
  \emph {et~al.}}]{zhong201812}%
  \BibitemOpen
  \bibfield  {author} {\bibinfo {author} {\bibfnamefont {H.-S.}\ \bibnamefont
  {Zhong}} \emph {et~al.},\ }\bibfield  {title} {\bibinfo {title} {12-photon
  entanglement and scalable scattershot boson sampling with optimal
  entangled-photon pairs from parametric down-conversion},\ }\href
  {https://doi.org/10.1103/PhysRevLett.121.250505} {\bibfield  {journal}
  {\bibinfo  {journal} {Phys. Rev. Lett.}\ }\textbf {\bibinfo {volume} {121}},\
  \bibinfo {pages} {250505} (\bibinfo {year} {2018})}\BibitemShut {NoStop}%
\bibitem [{\citenamefont {Zhong}\ \emph {et~al.}(2019)\citenamefont {Zhong}
  \emph {et~al.}}]{zhong2019experimental}%
  \BibitemOpen
  \bibfield  {author} {\bibinfo {author} {\bibfnamefont {H.-S.}\ \bibnamefont
  {Zhong}} \emph {et~al.},\ }\bibfield  {title} {\bibinfo {title} {Experimental
  {G}aussian boson sampling},\ }\href
  {https://doi.org/10.1016/j.scib.2019.04.007} {\bibfield  {journal} {\bibinfo
  {journal} {Science Bulletin}\ }\textbf {\bibinfo {volume} {64}},\ \bibinfo
  {pages} {511} (\bibinfo {year} {2019})}\BibitemShut {NoStop}%
\bibitem [{\citenamefont {Paesani}\ \emph {et~al.}(2019)\citenamefont
  {Paesani}, \citenamefont {Ding}, \citenamefont {Santagati}, \citenamefont
  {Chakhmakhchyan}, \citenamefont {Vigliar}, \citenamefont {Rottwitt},
  \citenamefont {Oxenl{\o}we}, \citenamefont {Wang}, \citenamefont {Thompson},\
  and\ \citenamefont {Laing}}]{paesani2019generation}%
  \BibitemOpen
  \bibfield  {author} {\bibinfo {author} {\bibfnamefont {S.}~\bibnamefont
  {Paesani}}, \bibinfo {author} {\bibfnamefont {Y.}~\bibnamefont {Ding}},
  \bibinfo {author} {\bibfnamefont {R.}~\bibnamefont {Santagati}}, \bibinfo
  {author} {\bibfnamefont {L.}~\bibnamefont {Chakhmakhchyan}}, \bibinfo
  {author} {\bibfnamefont {C.}~\bibnamefont {Vigliar}}, \bibinfo {author}
  {\bibfnamefont {K.}~\bibnamefont {Rottwitt}}, \bibinfo {author}
  {\bibfnamefont {L.~K.}\ \bibnamefont {Oxenl{\o}we}}, \bibinfo {author}
  {\bibfnamefont {J.}~\bibnamefont {Wang}}, \bibinfo {author} {\bibfnamefont
  {M.~G.}\ \bibnamefont {Thompson}},\ and\ \bibinfo {author} {\bibfnamefont
  {A.}~\bibnamefont {Laing}},\ }\bibfield  {title} {\bibinfo {title}
  {Generation and sampling of quantum states of light in a silicon chip},\
  }\href {https://doi.org/10.1038/s41567-019-0567-8} {\bibfield  {journal}
  {\bibinfo  {journal} {Nat. Phys.}\ }\textbf {\bibinfo {volume} {15}},\
  \bibinfo {pages} {925} (\bibinfo {year} {2019})}\BibitemShut {NoStop}%
\bibitem [{\citenamefont {He}\ \emph {et~al.}(2017)\citenamefont {He},
  \citenamefont {Ding}, \citenamefont {Su}, \citenamefont {Huang},
  \citenamefont {Qin}, \citenamefont {Wang}, \citenamefont {Unsleber},
  \citenamefont {Chen}, \citenamefont {Wang}, \citenamefont {He} \emph
  {et~al.}}]{he2017time}%
  \BibitemOpen
  \bibfield  {author} {\bibinfo {author} {\bibfnamefont {Y.}~\bibnamefont
  {He}}, \bibinfo {author} {\bibfnamefont {X.}~\bibnamefont {Ding}}, \bibinfo
  {author} {\bibfnamefont {Z.-E.}\ \bibnamefont {Su}}, \bibinfo {author}
  {\bibfnamefont {H.-L.}\ \bibnamefont {Huang}}, \bibinfo {author}
  {\bibfnamefont {J.}~\bibnamefont {Qin}}, \bibinfo {author} {\bibfnamefont
  {C.}~\bibnamefont {Wang}}, \bibinfo {author} {\bibfnamefont {S.}~\bibnamefont
  {Unsleber}}, \bibinfo {author} {\bibfnamefont {C.}~\bibnamefont {Chen}},
  \bibinfo {author} {\bibfnamefont {H.}~\bibnamefont {Wang}}, \bibinfo {author}
  {\bibfnamefont {Y.-M.}\ \bibnamefont {He}}, \emph {et~al.},\ }\bibfield
  {title} {\bibinfo {title} {Time-bin-encoded boson sampling with a
  single-photon device},\ }\href
  {https://doi.org/10.1103/PhysRevLett.118.190501} {\bibfield  {journal}
  {\bibinfo  {journal} {Phys. Rev. Lett.}\ }\textbf {\bibinfo {volume} {118}},\
  \bibinfo {pages} {190501} (\bibinfo {year} {2017})}\BibitemShut {NoStop}%
\bibitem [{\citenamefont {Loredo}\ \emph {et~al.}(2017)\citenamefont {Loredo},
  \citenamefont {Broome}, \citenamefont {Hilaire}, \citenamefont {Gazzano},
  \citenamefont {Sagnes}, \citenamefont {Lemaitre}, \citenamefont {Almeida},
  \citenamefont {Senellart},\ and\ \citenamefont {White}}]{loredo2017boson}%
  \BibitemOpen
  \bibfield  {author} {\bibinfo {author} {\bibfnamefont {J.~C.}\ \bibnamefont
  {Loredo}}, \bibinfo {author} {\bibfnamefont {M.~A.}\ \bibnamefont {Broome}},
  \bibinfo {author} {\bibfnamefont {P.}~\bibnamefont {Hilaire}}, \bibinfo
  {author} {\bibfnamefont {O.}~\bibnamefont {Gazzano}}, \bibinfo {author}
  {\bibfnamefont {I.}~\bibnamefont {Sagnes}}, \bibinfo {author} {\bibfnamefont
  {A.}~\bibnamefont {Lemaitre}}, \bibinfo {author} {\bibfnamefont {M.~P.}\
  \bibnamefont {Almeida}}, \bibinfo {author} {\bibfnamefont {P.}~\bibnamefont
  {Senellart}},\ and\ \bibinfo {author} {\bibfnamefont {A.~G.}\ \bibnamefont
  {White}},\ }\bibfield  {title} {\bibinfo {title} {Boson sampling with
  single-photon fock states from a bright solid-state source},\ }\href
  {https://doi.org/10.1103/PhysRevLett.118.130503} {\bibfield  {journal}
  {\bibinfo  {journal} {Phys. Rev. Lett.}\ }\textbf {\bibinfo {volume} {118}},\
  \bibinfo {pages} {130503} (\bibinfo {year} {2017})}\BibitemShut {NoStop}%
\bibitem [{\citenamefont {Wang}\ \emph {et~al.}(2017)\citenamefont {Wang},
  \citenamefont {He}, \citenamefont {Li}, \citenamefont {Su}, \citenamefont
  {Li}, \citenamefont {Huang}, \citenamefont {Ding}, \citenamefont {Chen},
  \citenamefont {Liu}, \citenamefont {Qin} \emph {et~al.}}]{wang2017high}%
  \BibitemOpen
  \bibfield  {author} {\bibinfo {author} {\bibfnamefont {H.}~\bibnamefont
  {Wang}}, \bibinfo {author} {\bibfnamefont {Y.}~\bibnamefont {He}}, \bibinfo
  {author} {\bibfnamefont {Y.-H.}\ \bibnamefont {Li}}, \bibinfo {author}
  {\bibfnamefont {Z.-E.}\ \bibnamefont {Su}}, \bibinfo {author} {\bibfnamefont
  {B.}~\bibnamefont {Li}}, \bibinfo {author} {\bibfnamefont {H.-L.}\
  \bibnamefont {Huang}}, \bibinfo {author} {\bibfnamefont {X.}~\bibnamefont
  {Ding}}, \bibinfo {author} {\bibfnamefont {M.-C.}\ \bibnamefont {Chen}},
  \bibinfo {author} {\bibfnamefont {C.}~\bibnamefont {Liu}}, \bibinfo {author}
  {\bibfnamefont {J.}~\bibnamefont {Qin}}, \emph {et~al.},\ }\bibfield  {title}
  {\bibinfo {title} {High-efficiency multiphoton boson sampling},\ }\href
  {https://doi.org/10.1038/nphoton.2017.63} {\bibfield  {journal} {\bibinfo
  {journal} {Nat. Photonics}\ }\textbf {\bibinfo {volume} {11}},\ \bibinfo
  {pages} {361} (\bibinfo {year} {2017})}\BibitemShut {NoStop}%
\bibitem [{\citenamefont {Wang}\ \emph {et~al.}(2018)\citenamefont {Wang},
  \citenamefont {Li}, \citenamefont {Jiang}, \citenamefont {He}, \citenamefont
  {Li}, \citenamefont {Ding}, \citenamefont {Chen}, \citenamefont {Qin},
  \citenamefont {Peng}, \citenamefont {Schneider} \emph
  {et~al.}}]{wang2018toward}%
  \BibitemOpen
  \bibfield  {author} {\bibinfo {author} {\bibfnamefont {H.}~\bibnamefont
  {Wang}}, \bibinfo {author} {\bibfnamefont {W.}~\bibnamefont {Li}}, \bibinfo
  {author} {\bibfnamefont {X.}~\bibnamefont {Jiang}}, \bibinfo {author}
  {\bibfnamefont {Y.-M.}\ \bibnamefont {He}}, \bibinfo {author} {\bibfnamefont
  {Y.-H.}\ \bibnamefont {Li}}, \bibinfo {author} {\bibfnamefont
  {X.}~\bibnamefont {Ding}}, \bibinfo {author} {\bibfnamefont {M.-C.}\
  \bibnamefont {Chen}}, \bibinfo {author} {\bibfnamefont {J.}~\bibnamefont
  {Qin}}, \bibinfo {author} {\bibfnamefont {C.-Z.}\ \bibnamefont {Peng}},
  \bibinfo {author} {\bibfnamefont {C.}~\bibnamefont {Schneider}}, \emph
  {et~al.},\ }\bibfield  {title} {\bibinfo {title} {Toward scalable boson
  sampling with photon loss},\ }\href
  {https://doi.org/10.1103/PhysRevLett.120.230502} {\bibfield  {journal}
  {\bibinfo  {journal} {Phys. Rev. Lett.}\ }\textbf {\bibinfo {volume} {120}},\
  \bibinfo {pages} {230502} (\bibinfo {year} {2018})}\BibitemShut {NoStop}%
\bibitem [{\citenamefont {Wang}\ \emph {et~al.}(2019)\citenamefont {Wang},
  \citenamefont {Qin}, \citenamefont {Ding}, \citenamefont {Chen},
  \citenamefont {Chen}, \citenamefont {You}, \citenamefont {He}, \citenamefont
  {Jiang}, \citenamefont {You}, \citenamefont {Wang} \emph
  {et~al.}}]{wang2019boson}%
  \BibitemOpen
  \bibfield  {author} {\bibinfo {author} {\bibfnamefont {H.}~\bibnamefont
  {Wang}}, \bibinfo {author} {\bibfnamefont {J.}~\bibnamefont {Qin}}, \bibinfo
  {author} {\bibfnamefont {X.}~\bibnamefont {Ding}}, \bibinfo {author}
  {\bibfnamefont {M.-C.}\ \bibnamefont {Chen}}, \bibinfo {author}
  {\bibfnamefont {S.}~\bibnamefont {Chen}}, \bibinfo {author} {\bibfnamefont
  {X.}~\bibnamefont {You}}, \bibinfo {author} {\bibfnamefont {Y.-M.}\
  \bibnamefont {He}}, \bibinfo {author} {\bibfnamefont {X.}~\bibnamefont
  {Jiang}}, \bibinfo {author} {\bibfnamefont {L.}~\bibnamefont {You}}, \bibinfo
  {author} {\bibfnamefont {Z.}~\bibnamefont {Wang}}, \emph {et~al.},\
  }\bibfield  {title} {\bibinfo {title} {Boson sampling with 20 input photons
  and a 60-mode interferometer in a ${10^{14}}$-dimensional hilbert space},\
  }\href {https://doi.org/10.1103/PhysRevLett.123.250503} {\bibfield  {journal}
  {\bibinfo  {journal} {Phys. Rev. Lett.}\ }\textbf {\bibinfo {volume} {123}},\
  \bibinfo {pages} {250503} (\bibinfo {year} {2019})}\BibitemShut {NoStop}%
\bibitem [{\citenamefont {Zhong}\ \emph {et~al.}(2020)\citenamefont {Zhong}
  \emph {et~al.}}]{zhong2020quantum}%
  \BibitemOpen
  \bibfield  {author} {\bibinfo {author} {\bibfnamefont {H.-S.}\ \bibnamefont
  {Zhong}} \emph {et~al.},\ }\bibfield  {title} {\bibinfo {title} {Quantum
  computational advantage using photons},\ }\bibfield  {journal} {\bibinfo
  {journal} {Science}\ }\href {https://doi.org/10.1126/science.abe8770}
  {10.1126/science.abe8770} (\bibinfo {year} {2020})\BibitemShut {NoStop}%
\bibitem [{\citenamefont {Zhong}\ \emph {et~al.}(2021)\citenamefont {Zhong}
  \emph {et~al.}}]{zhong2021phase}%
  \BibitemOpen
  \bibfield  {author} {\bibinfo {author} {\bibfnamefont {H.-S.}\ \bibnamefont
  {Zhong}} \emph {et~al.},\ }\bibfield  {title} {\bibinfo {title}
  {Phase-programmable gaussian boson sampling using stimulated squeezed
  light},\ }\href {https://doi.org/10.1103/PhysRevLett.127.180502} {\bibfield
  {journal} {\bibinfo  {journal} {Phys. Rev. Lett.}\ }\textbf {\bibinfo
  {volume} {127}},\ \bibinfo {pages} {180502} (\bibinfo {year}
  {2021})}\BibitemShut {NoStop}%
\bibitem [{\citenamefont {Madsen}\ \emph {et~al.}(2022)\citenamefont {Madsen}
  \emph {et~al.}}]{madsen2022quantum}%
  \BibitemOpen
  \bibfield  {author} {\bibinfo {author} {\bibfnamefont {L.~S.}\ \bibnamefont
  {Madsen}} \emph {et~al.},\ }\bibfield  {title} {\bibinfo {title} {Quantum
  computational advantage with a programmable photonic processor},\ }\href
  {https://doi.org/10.1038/s41586-022-04725-x} {\bibfield  {journal} {\bibinfo
  {journal} {Nature}\ }\textbf {\bibinfo {volume} {606}},\ \bibinfo {pages}
  {75} (\bibinfo {year} {2022})}\BibitemShut {NoStop}%
\bibitem [{\citenamefont {Aharonov}\ \emph {et~al.}(1996)\citenamefont
  {Aharonov}, \citenamefont {Ben-Or}, \citenamefont {Impagliazzo},\ and\
  \citenamefont {Nisan}}]{aharonov1996limitations}%
  \BibitemOpen
  \bibfield  {author} {\bibinfo {author} {\bibfnamefont {D.}~\bibnamefont
  {Aharonov}}, \bibinfo {author} {\bibfnamefont {M.}~\bibnamefont {Ben-Or}},
  \bibinfo {author} {\bibfnamefont {R.}~\bibnamefont {Impagliazzo}},\ and\
  \bibinfo {author} {\bibfnamefont {N.}~\bibnamefont {Nisan}},\ }\bibfield
  {title} {\bibinfo {title} {Limitations of noisy reversible computation},\
  }\href@noop {} {\bibfield  {journal} {\bibinfo  {journal} {arXiv preprint
  quant-ph/9611028}\ } (\bibinfo {year} {1996})}\BibitemShut {NoStop}%
\bibitem [{\citenamefont {Noh}\ \emph {et~al.}(2020)\citenamefont {Noh},
  \citenamefont {Jiang},\ and\ \citenamefont {Fefferman}}]{noh2020efficient}%
  \BibitemOpen
  \bibfield  {author} {\bibinfo {author} {\bibfnamefont {K.}~\bibnamefont
  {Noh}}, \bibinfo {author} {\bibfnamefont {L.}~\bibnamefont {Jiang}},\ and\
  \bibinfo {author} {\bibfnamefont {B.}~\bibnamefont {Fefferman}},\ }\bibfield
  {title} {\bibinfo {title} {Efficient classical simulation of noisy random
  quantum circuits in one dimension},\ }\href
  {https://doi.org/10.22331/q-2020-09-11-318} {\bibfield  {journal} {\bibinfo
  {journal} {{Quantum}}\ }\textbf {\bibinfo {volume} {4}},\ \bibinfo {pages}
  {318} (\bibinfo {year} {2020})}\BibitemShut {NoStop}%
\bibitem [{\citenamefont {Zhang}\ \emph {et~al.}(2022)\citenamefont {Zhang},
  \citenamefont {Wang}, \citenamefont {Dong}, \citenamefont {Zhang},
  \citenamefont {Han},\ and\ \citenamefont {He}}]{zhang2022entanglement}%
  \BibitemOpen
  \bibfield  {author} {\bibinfo {author} {\bibfnamefont {M.}~\bibnamefont
  {Zhang}}, \bibinfo {author} {\bibfnamefont {C.}~\bibnamefont {Wang}},
  \bibinfo {author} {\bibfnamefont {S.}~\bibnamefont {Dong}}, \bibinfo {author}
  {\bibfnamefont {H.}~\bibnamefont {Zhang}}, \bibinfo {author} {\bibfnamefont
  {Y.}~\bibnamefont {Han}},\ and\ \bibinfo {author} {\bibfnamefont
  {L.}~\bibnamefont {He}},\ }\bibfield  {title} {\bibinfo {title} {Entanglement
  entropy scaling of noisy random quantum circuits in two dimensions},\ }\href
  {https://doi.org/10.1103/PhysRevA.106.052430} {\bibfield  {journal} {\bibinfo
   {journal} {Phys. Rev. A}\ }\textbf {\bibinfo {volume} {106}},\ \bibinfo
  {pages} {052430} (\bibinfo {year} {2022})}\BibitemShut {NoStop}%
\bibitem [{\citenamefont {Oh}\ \emph {et~al.}(2021)\citenamefont {Oh},
  \citenamefont {Noh}, \citenamefont {Fefferman},\ and\ \citenamefont
  {Jiang}}]{oh2021classical}%
  \BibitemOpen
  \bibfield  {author} {\bibinfo {author} {\bibfnamefont {C.}~\bibnamefont
  {Oh}}, \bibinfo {author} {\bibfnamefont {K.}~\bibnamefont {Noh}}, \bibinfo
  {author} {\bibfnamefont {B.}~\bibnamefont {Fefferman}},\ and\ \bibinfo
  {author} {\bibfnamefont {L.}~\bibnamefont {Jiang}},\ }\bibfield  {title}
  {\bibinfo {title} {Classical simulation of lossy boson sampling using matrix
  product operators},\ }\href {https://doi.org/10.1103/PhysRevA.104.022407}
  {\bibfield  {journal} {\bibinfo  {journal} {Phys. Rev. A}\ }\textbf {\bibinfo
  {volume} {104}},\ \bibinfo {pages} {022407} (\bibinfo {year}
  {2021})}\BibitemShut {NoStop}%
\bibitem [{\citenamefont {Arute}\ \emph {et~al.}(2019)\citenamefont {Arute}
  \emph {et~al.}}]{arute2019quantum}%
  \BibitemOpen
  \bibfield  {author} {\bibinfo {author} {\bibfnamefont {F.}~\bibnamefont
  {Arute}} \emph {et~al.},\ }\bibfield  {title} {\bibinfo {title} {Quantum
  supremacy using a programmable superconducting processor},\ }\href@noop {}
  {\bibfield  {journal} {\bibinfo  {journal} {Nature}\ }\textbf {\bibinfo
  {volume} {574}},\ \bibinfo {pages} {505} (\bibinfo {year}
  {2019})}\BibitemShut {NoStop}%
\bibitem [{\citenamefont {Aharonov}\ \emph {et~al.}(2022)\citenamefont
  {Aharonov}, \citenamefont {Gao}, \citenamefont {Landau}, \citenamefont
  {Liu},\ and\ \citenamefont {Vazirani}}]{aharonov2022polynomial}%
  \BibitemOpen
  \bibfield  {author} {\bibinfo {author} {\bibfnamefont {D.}~\bibnamefont
  {Aharonov}}, \bibinfo {author} {\bibfnamefont {X.}~\bibnamefont {Gao}},
  \bibinfo {author} {\bibfnamefont {Z.}~\bibnamefont {Landau}}, \bibinfo
  {author} {\bibfnamefont {Y.}~\bibnamefont {Liu}},\ and\ \bibinfo {author}
  {\bibfnamefont {U.}~\bibnamefont {Vazirani}},\ }\bibfield  {title} {\bibinfo
  {title} {A polynomial-time classical algorithm for noisy random circuit
  sampling},\ }\href@noop {} {\bibfield  {journal} {\bibinfo  {journal} {arXiv
  preprint arXiv:2211.03999}\ } (\bibinfo {year} {2022})}\BibitemShut {NoStop}%
\bibitem [{\citenamefont {Oh}\ \emph {et~al.}(2023{\natexlab{a}})\citenamefont
  {Oh}, \citenamefont {Jiang},\ and\ \citenamefont
  {Fefferman}}]{oh2023classical}%
  \BibitemOpen
  \bibfield  {author} {\bibinfo {author} {\bibfnamefont {C.}~\bibnamefont
  {Oh}}, \bibinfo {author} {\bibfnamefont {L.}~\bibnamefont {Jiang}},\ and\
  \bibinfo {author} {\bibfnamefont {B.}~\bibnamefont {Fefferman}},\ }\bibfield
  {title} {\bibinfo {title} {On classical simulation algorithms for noisy boson
  sampling},\ }\href@noop {} {\bibfield  {journal} {\bibinfo  {journal} {arXiv
  preprint arXiv:2301.11532}\ } (\bibinfo {year}
  {2023}{\natexlab{a}})}\BibitemShut {NoStop}%
\bibitem [{\citenamefont {Oszmaniec}\ and\ \citenamefont
  {Brod}(2018)}]{oszmaniec2018classical}%
  \BibitemOpen
  \bibfield  {author} {\bibinfo {author} {\bibfnamefont {M.}~\bibnamefont
  {Oszmaniec}}\ and\ \bibinfo {author} {\bibfnamefont {D.~J.}\ \bibnamefont
  {Brod}},\ }\bibfield  {title} {\bibinfo {title} {Classical simulation of
  photonic linear optics with lost particles},\ }\href
  {https://doi.org/10.1088/1367-2630/aadfa8} {\bibfield  {journal} {\bibinfo
  {journal} {New J. Phys.}\ }\textbf {\bibinfo {volume} {20}},\ \bibinfo
  {pages} {092002} (\bibinfo {year} {2018})}\BibitemShut {NoStop}%
\bibitem [{\citenamefont {Garc{\'\i}a-Patr{\'o}n}\ \emph
  {et~al.}(2019)\citenamefont {Garc{\'\i}a-Patr{\'o}n}, \citenamefont
  {Renema},\ and\ \citenamefont {Shchesnovich}}]{garcia2019simulating}%
  \BibitemOpen
  \bibfield  {author} {\bibinfo {author} {\bibfnamefont {R.}~\bibnamefont
  {Garc{\'\i}a-Patr{\'o}n}}, \bibinfo {author} {\bibfnamefont {J.~J.}\
  \bibnamefont {Renema}},\ and\ \bibinfo {author} {\bibfnamefont
  {V.}~\bibnamefont {Shchesnovich}},\ }\bibfield  {title} {\bibinfo {title}
  {Simulating boson sampling in lossy architectures},\ }\href
  {https://doi.org/10.22331/q-2019-08-05-169} {\bibfield  {journal} {\bibinfo
  {journal} {Quantum}\ }\textbf {\bibinfo {volume} {3}},\ \bibinfo {pages}
  {169} (\bibinfo {year} {2019})}\BibitemShut {NoStop}%
\bibitem [{\citenamefont {Renema}\ \emph
  {et~al.}(2018{\natexlab{a}})\citenamefont {Renema}, \citenamefont
  {Shchesnovich},\ and\ \citenamefont {Garcia-Patron}}]{renema2018classical}%
  \BibitemOpen
  \bibfield  {author} {\bibinfo {author} {\bibfnamefont {J.}~\bibnamefont
  {Renema}}, \bibinfo {author} {\bibfnamefont {V.}~\bibnamefont
  {Shchesnovich}},\ and\ \bibinfo {author} {\bibfnamefont {R.}~\bibnamefont
  {Garcia-Patron}},\ }\bibfield  {title} {\bibinfo {title} {Classical
  simulability of noisy boson sampling},\ }\href@noop {} {\bibfield  {journal}
  {\bibinfo  {journal} {arXiv preprint arXiv:1809.01953}\ } (\bibinfo {year}
  {2018}{\natexlab{a}})}\BibitemShut {NoStop}%
\bibitem [{\citenamefont {Qi}\ \emph {et~al.}(2020)\citenamefont {Qi},
  \citenamefont {Brod}, \citenamefont {Quesada},\ and\ \citenamefont
  {Garc\'{\i}a-Patr\'on}}]{qi2020regimes}%
  \BibitemOpen
  \bibfield  {author} {\bibinfo {author} {\bibfnamefont {H.}~\bibnamefont
  {Qi}}, \bibinfo {author} {\bibfnamefont {D.~J.}\ \bibnamefont {Brod}},
  \bibinfo {author} {\bibfnamefont {N.}~\bibnamefont {Quesada}},\ and\ \bibinfo
  {author} {\bibfnamefont {R.}~\bibnamefont {Garc\'{\i}a-Patr\'on}},\
  }\bibfield  {title} {\bibinfo {title} {Regimes of classical simulability for
  noisy {G}aussian boson sampling},\ }\href
  {https://doi.org/10.1103/PhysRevLett.124.100502} {\bibfield  {journal}
  {\bibinfo  {journal} {Phys. Rev. Lett.}\ }\textbf {\bibinfo {volume} {124}},\
  \bibinfo {pages} {100502} (\bibinfo {year} {2020})}\BibitemShut {NoStop}%
\bibitem [{\citenamefont {Tichy}(2015)}]{tichy2015sampling}%
  \BibitemOpen
  \bibfield  {author} {\bibinfo {author} {\bibfnamefont {M.~C.}\ \bibnamefont
  {Tichy}},\ }\bibfield  {title} {\bibinfo {title} {Sampling of partially
  distinguishable bosons and the relation to the multidimensional permanent},\
  }\href@noop {} {\bibfield  {journal} {\bibinfo  {journal} {Physical Review
  A}\ }\textbf {\bibinfo {volume} {91}},\ \bibinfo {pages} {022316} (\bibinfo
  {year} {2015})}\BibitemShut {NoStop}%
\bibitem [{\citenamefont {Renema}\ \emph
  {et~al.}(2018{\natexlab{b}})\citenamefont {Renema}, \citenamefont {Menssen},
  \citenamefont {Clements}, \citenamefont {Triginer}, \citenamefont
  {Kolthammer},\ and\ \citenamefont {Walmsley}}]{renema2018efficient}%
  \BibitemOpen
  \bibfield  {author} {\bibinfo {author} {\bibfnamefont {J.~J.}\ \bibnamefont
  {Renema}}, \bibinfo {author} {\bibfnamefont {A.}~\bibnamefont {Menssen}},
  \bibinfo {author} {\bibfnamefont {W.~R.}\ \bibnamefont {Clements}}, \bibinfo
  {author} {\bibfnamefont {G.}~\bibnamefont {Triginer}}, \bibinfo {author}
  {\bibfnamefont {W.~S.}\ \bibnamefont {Kolthammer}},\ and\ \bibinfo {author}
  {\bibfnamefont {I.~A.}\ \bibnamefont {Walmsley}},\ }\bibfield  {title}
  {\bibinfo {title} {Efficient classical algorithm for boson sampling with
  partially distinguishable photons},\ }\href@noop {} {\bibfield  {journal}
  {\bibinfo  {journal} {Physical review letters}\ }\textbf {\bibinfo {volume}
  {120}},\ \bibinfo {pages} {220502} (\bibinfo {year}
  {2018}{\natexlab{b}})}\BibitemShut {NoStop}%
\bibitem [{\citenamefont {Shchesnovich}(2019)}]{shchesnovich2019noise}%
  \BibitemOpen
  \bibfield  {author} {\bibinfo {author} {\bibfnamefont {V.~S.}\ \bibnamefont
  {Shchesnovich}},\ }\bibfield  {title} {\bibinfo {title} {Noise in boson
  sampling and the threshold of efficient classical simulatability},\
  }\href@noop {} {\bibfield  {journal} {\bibinfo  {journal} {Physical Review
  A}\ }\textbf {\bibinfo {volume} {100}},\ \bibinfo {pages} {012340} (\bibinfo
  {year} {2019})}\BibitemShut {NoStop}%
\bibitem [{\citenamefont {Moylett}\ \emph {et~al.}(2019)\citenamefont
  {Moylett}, \citenamefont {Garc{\'\i}a-Patr{\'o}n}, \citenamefont {Renema},\
  and\ \citenamefont {Turner}}]{moylett2019classically}%
  \BibitemOpen
  \bibfield  {author} {\bibinfo {author} {\bibfnamefont {A.~E.}\ \bibnamefont
  {Moylett}}, \bibinfo {author} {\bibfnamefont {R.}~\bibnamefont
  {Garc{\'\i}a-Patr{\'o}n}}, \bibinfo {author} {\bibfnamefont {J.~J.}\
  \bibnamefont {Renema}},\ and\ \bibinfo {author} {\bibfnamefont {P.~S.}\
  \bibnamefont {Turner}},\ }\bibfield  {title} {\bibinfo {title} {Classically
  simulating near-term partially-distinguishable and lossy boson sampling},\
  }\href@noop {} {\bibfield  {journal} {\bibinfo  {journal} {Quantum Science
  and Technology}\ }\textbf {\bibinfo {volume} {5}},\ \bibinfo {pages} {015001}
  (\bibinfo {year} {2019})}\BibitemShut {NoStop}%
\bibitem [{\citenamefont {Martínez-Cifuentes}\ \emph
  {et~al.}(2023)\citenamefont {Martínez-Cifuentes}, \citenamefont
  {Fonseca-Romero},\ and\ \citenamefont {Quesada}}]{cifuentes2023classical}%
  \BibitemOpen
  \bibfield  {author} {\bibinfo {author} {\bibfnamefont {J.}~\bibnamefont
  {Martínez-Cifuentes}}, \bibinfo {author} {\bibfnamefont {K.~M.}\
  \bibnamefont {Fonseca-Romero}},\ and\ \bibinfo {author} {\bibfnamefont
  {N.}~\bibnamefont {Quesada}},\ }\bibfield  {title} {\bibinfo {title}
  {Classical models may be a better explanation of the jiuzhang 1.0 gaussian
  boson sampler than its targeted squeezed light model},\ }\href
  {https://doi.org/10.22331/q-2023-08-08-1076} {\bibfield  {journal} {\bibinfo
  {journal} {Quantum}\ }\textbf {\bibinfo {volume} {7}},\ \bibinfo {pages}
  {1076} (\bibinfo {year} {2023})}\BibitemShut {NoStop}%
\bibitem [{\citenamefont {Lund}\ \emph {et~al.}(2014)\citenamefont {Lund},
  \citenamefont {Laing}, \citenamefont {Rahimi-Keshari}, \citenamefont
  {Rudolph}, \citenamefont {O'Brien},\ and\ \citenamefont
  {Ralph}}]{lund2014boson}%
  \BibitemOpen
  \bibfield  {author} {\bibinfo {author} {\bibfnamefont {A.~P.}\ \bibnamefont
  {Lund}}, \bibinfo {author} {\bibfnamefont {A.}~\bibnamefont {Laing}},
  \bibinfo {author} {\bibfnamefont {S.}~\bibnamefont {Rahimi-Keshari}},
  \bibinfo {author} {\bibfnamefont {T.}~\bibnamefont {Rudolph}}, \bibinfo
  {author} {\bibfnamefont {J.~L.}\ \bibnamefont {O'Brien}},\ and\ \bibinfo
  {author} {\bibfnamefont {T.~C.}\ \bibnamefont {Ralph}},\ }\bibfield  {title}
  {\bibinfo {title} {Boson sampling from a gaussian state},\ }\href
  {https://doi.org/10.1103/PhysRevLett.113.100502} {\bibfield  {journal}
  {\bibinfo  {journal} {Phys. Rev. Lett.}\ }\textbf {\bibinfo {volume} {113}},\
  \bibinfo {pages} {100502} (\bibinfo {year} {2014})}\BibitemShut {NoStop}%
\bibitem [{\citenamefont {Barkhofen}\ \emph {et~al.}(2017)\citenamefont
  {Barkhofen}, \citenamefont {Bartley}, \citenamefont {Sansoni}, \citenamefont
  {Kruse}, \citenamefont {Hamilton}, \citenamefont {Jex},\ and\ \citenamefont
  {Silberhorn}}]{barkhofen2017driven}%
  \BibitemOpen
  \bibfield  {author} {\bibinfo {author} {\bibfnamefont {S.}~\bibnamefont
  {Barkhofen}}, \bibinfo {author} {\bibfnamefont {T.~J.}\ \bibnamefont
  {Bartley}}, \bibinfo {author} {\bibfnamefont {L.}~\bibnamefont {Sansoni}},
  \bibinfo {author} {\bibfnamefont {R.}~\bibnamefont {Kruse}}, \bibinfo
  {author} {\bibfnamefont {C.~S.}\ \bibnamefont {Hamilton}}, \bibinfo {author}
  {\bibfnamefont {I.}~\bibnamefont {Jex}},\ and\ \bibinfo {author}
  {\bibfnamefont {C.}~\bibnamefont {Silberhorn}},\ }\bibfield  {title}
  {\bibinfo {title} {Driven boson sampling},\ }\href
  {https://doi.org/10.1103/PhysRevLett.118.020502} {\bibfield  {journal}
  {\bibinfo  {journal} {Phys. Rev. Lett.}\ }\textbf {\bibinfo {volume} {118}},\
  \bibinfo {pages} {020502} (\bibinfo {year} {2017})}\BibitemShut {NoStop}%
\bibitem [{\citenamefont {Chakhmakhchyan}\ and\ \citenamefont
  {Cerf}(2017)}]{chakmakhchyan2017boson}%
  \BibitemOpen
  \bibfield  {author} {\bibinfo {author} {\bibfnamefont {L.}~\bibnamefont
  {Chakhmakhchyan}}\ and\ \bibinfo {author} {\bibfnamefont {N.~J.}\
  \bibnamefont {Cerf}},\ }\bibfield  {title} {\bibinfo {title} {Boson sampling
  with gaussian measurements},\ }\href
  {https://doi.org/10.1103/PhysRevA.96.032326} {\bibfield  {journal} {\bibinfo
  {journal} {Phys. Rev. A}\ }\textbf {\bibinfo {volume} {96}},\ \bibinfo
  {pages} {032326} (\bibinfo {year} {2017})}\BibitemShut {NoStop}%
\bibitem [{\citenamefont {Hamilton}\ \emph {et~al.}(2017)\citenamefont
  {Hamilton}, \citenamefont {Kruse}, \citenamefont {Sansoni}, \citenamefont
  {Barkhofen}, \citenamefont {Silberhorn},\ and\ \citenamefont
  {Jex}}]{hamilton2017gaussian}%
  \BibitemOpen
  \bibfield  {author} {\bibinfo {author} {\bibfnamefont {C.~S.}\ \bibnamefont
  {Hamilton}}, \bibinfo {author} {\bibfnamefont {R.}~\bibnamefont {Kruse}},
  \bibinfo {author} {\bibfnamefont {L.}~\bibnamefont {Sansoni}}, \bibinfo
  {author} {\bibfnamefont {S.}~\bibnamefont {Barkhofen}}, \bibinfo {author}
  {\bibfnamefont {C.}~\bibnamefont {Silberhorn}},\ and\ \bibinfo {author}
  {\bibfnamefont {I.}~\bibnamefont {Jex}},\ }\bibfield  {title} {\bibinfo
  {title} {Gaussian boson sampling},\ }\href
  {https://doi.org/10.1103/PhysRevLett.119.170501} {\bibfield  {journal}
  {\bibinfo  {journal} {Phys. Rev. Lett.}\ }\textbf {\bibinfo {volume} {119}},\
  \bibinfo {pages} {170501} (\bibinfo {year} {2017})}\BibitemShut {NoStop}%
\bibitem [{\citenamefont {Kruse}\ \emph {et~al.}(2019)\citenamefont {Kruse},
  \citenamefont {Hamilton}, \citenamefont {Sansoni}, \citenamefont {Barkhofen},
  \citenamefont {Silberhorn},\ and\ \citenamefont {Jex}}]{kruse2019detailed}%
  \BibitemOpen
  \bibfield  {author} {\bibinfo {author} {\bibfnamefont {R.}~\bibnamefont
  {Kruse}}, \bibinfo {author} {\bibfnamefont {C.~S.}\ \bibnamefont {Hamilton}},
  \bibinfo {author} {\bibfnamefont {L.}~\bibnamefont {Sansoni}}, \bibinfo
  {author} {\bibfnamefont {S.}~\bibnamefont {Barkhofen}}, \bibinfo {author}
  {\bibfnamefont {C.}~\bibnamefont {Silberhorn}},\ and\ \bibinfo {author}
  {\bibfnamefont {I.}~\bibnamefont {Jex}},\ }\bibfield  {title} {\bibinfo
  {title} {Detailed study of gaussian boson sampling},\ }\href
  {https://doi.org/10.1103/PhysRevA.100.032326} {\bibfield  {journal} {\bibinfo
   {journal} {Phys. Rev. A}\ }\textbf {\bibinfo {volume} {100}},\ \bibinfo
  {pages} {032326} (\bibinfo {year} {2019})}\BibitemShut {NoStop}%
\bibitem [{\citenamefont {Deng}\ \emph {et~al.}(2023)\citenamefont {Deng} \emph
  {et~al.}}]{deng2023gaussian}%
  \BibitemOpen
  \bibfield  {author} {\bibinfo {author} {\bibfnamefont {Y.-H.}\ \bibnamefont
  {Deng}} \emph {et~al.},\ }\bibfield  {title} {\bibinfo {title} {Gaussian
  boson sampling with pseudo-photon-number resolving detectors and quantum
  computational advantage},\ }\href@noop {} {\bibfield  {journal} {\bibinfo
  {journal} {arXiv preprint arXiv:2304.12240}\ } (\bibinfo {year}
  {2023})}\BibitemShut {NoStop}%
\bibitem [{\citenamefont {Liu}\ \emph {et~al.}(2023)\citenamefont {Liu},
  \citenamefont {Oh}, \citenamefont {Liu}, \citenamefont {Jiang},\ and\
  \citenamefont {Alexeev}}]{liu2023supercomputing}%
  \BibitemOpen
  \bibfield  {author} {\bibinfo {author} {\bibfnamefont {M.}~\bibnamefont
  {Liu}}, \bibinfo {author} {\bibfnamefont {C.}~\bibnamefont {Oh}}, \bibinfo
  {author} {\bibfnamefont {J.}~\bibnamefont {Liu}}, \bibinfo {author}
  {\bibfnamefont {L.}~\bibnamefont {Jiang}},\ and\ \bibinfo {author}
  {\bibfnamefont {Y.}~\bibnamefont {Alexeev}},\ }\bibfield  {title} {\bibinfo
  {title} {Supercomputing tensor networks for u(1) symmetric quantum many-body
  systems},\ }\href@noop {} {\bibfield  {journal} {\bibinfo  {journal} {arXiv
  preprint arXiv:2303.11409}\ } (\bibinfo {year} {2023})}\BibitemShut {NoStop}%
\bibitem [{\citenamefont {Quesada}\ and\ \citenamefont
  {Arrazola}(2020)}]{quesada2020exact}%
  \BibitemOpen
  \bibfield  {author} {\bibinfo {author} {\bibfnamefont {N.}~\bibnamefont
  {Quesada}}\ and\ \bibinfo {author} {\bibfnamefont {J.~M.}\ \bibnamefont
  {Arrazola}},\ }\bibfield  {title} {\bibinfo {title} {Exact simulation of
  gaussian boson sampling in polynomial space and exponential time},\ }\href
  {https://doi.org/10.1103/PhysRevResearch.2.023005} {\bibfield  {journal}
  {\bibinfo  {journal} {Phys. Rev. Res.}\ }\textbf {\bibinfo {volume} {2}},\
  \bibinfo {pages} {023005} (\bibinfo {year} {2020})}\BibitemShut {NoStop}%
\bibitem [{\citenamefont {Bulmer}\ \emph {et~al.}(2022)\citenamefont {Bulmer},
  \citenamefont {Bell}, \citenamefont {Chadwick}, \citenamefont {Jones},
  \citenamefont {Moise}, \citenamefont {Rigazzi}, \citenamefont {Thorbecke},
  \citenamefont {Haus}, \citenamefont {Vaerenbergh}, \citenamefont {Patel},
  \citenamefont {Walmsley},\ and\ \citenamefont {Laing}}]{bulmer2022boundary}%
  \BibitemOpen
  \bibfield  {author} {\bibinfo {author} {\bibfnamefont {J.~F.~F.}\
  \bibnamefont {Bulmer}}, \bibinfo {author} {\bibfnamefont {B.~A.}\
  \bibnamefont {Bell}}, \bibinfo {author} {\bibfnamefont {R.~S.}\ \bibnamefont
  {Chadwick}}, \bibinfo {author} {\bibfnamefont {A.~E.}\ \bibnamefont {Jones}},
  \bibinfo {author} {\bibfnamefont {D.}~\bibnamefont {Moise}}, \bibinfo
  {author} {\bibfnamefont {A.}~\bibnamefont {Rigazzi}}, \bibinfo {author}
  {\bibfnamefont {J.}~\bibnamefont {Thorbecke}}, \bibinfo {author}
  {\bibfnamefont {U.-U.}\ \bibnamefont {Haus}}, \bibinfo {author}
  {\bibfnamefont {T.~V.}\ \bibnamefont {Vaerenbergh}}, \bibinfo {author}
  {\bibfnamefont {R.~B.}\ \bibnamefont {Patel}}, \bibinfo {author}
  {\bibfnamefont {I.~A.}\ \bibnamefont {Walmsley}},\ and\ \bibinfo {author}
  {\bibfnamefont {A.}~\bibnamefont {Laing}},\ }\bibfield  {title} {\bibinfo
  {title} {The boundary for quantum advantage in gaussian boson sampling},\
  }\href {https://doi.org/10.1126/sciadv.abl9236} {\bibfield  {journal}
  {\bibinfo  {journal} {Science Advances}\ }\textbf {\bibinfo {volume} {8}},\
  \bibinfo {pages} {eabl9236} (\bibinfo {year} {2022})}\BibitemShut {NoStop}%
\bibitem [{\citenamefont {Quesada}\ \emph {et~al.}(2022)\citenamefont
  {Quesada}, \citenamefont {Chadwick}, \citenamefont {Bell}, \citenamefont
  {Arrazola}, \citenamefont {Vincent}, \citenamefont {Qi},\ and\ \citenamefont
  {Garc\'{\i}a\ensuremath{-}Patr\'on}}]{quesada2022quadratic}%
  \BibitemOpen
  \bibfield  {author} {\bibinfo {author} {\bibfnamefont {N.}~\bibnamefont
  {Quesada}}, \bibinfo {author} {\bibfnamefont {R.~S.}\ \bibnamefont
  {Chadwick}}, \bibinfo {author} {\bibfnamefont {B.~A.}\ \bibnamefont {Bell}},
  \bibinfo {author} {\bibfnamefont {J.~M.}\ \bibnamefont {Arrazola}}, \bibinfo
  {author} {\bibfnamefont {T.}~\bibnamefont {Vincent}}, \bibinfo {author}
  {\bibfnamefont {H.}~\bibnamefont {Qi}},\ and\ \bibinfo {author}
  {\bibfnamefont {R.}~\bibnamefont {Garc\'{\i}a\ensuremath{-}Patr\'on}},\
  }\bibfield  {title} {\bibinfo {title} {Quadratic speed-up for simulating
  gaussian boson sampling},\ }\href
  {https://doi.org/10.1103/PRXQuantum.3.010306} {\bibfield  {journal} {\bibinfo
   {journal} {PRX Quantum}\ }\textbf {\bibinfo {volume} {3}},\ \bibinfo {pages}
  {010306} (\bibinfo {year} {2022})}\BibitemShut {NoStop}%
\bibitem [{\citenamefont {Villalonga}\ \emph {et~al.}(2021)\citenamefont
  {Villalonga}, \citenamefont {Niu}, \citenamefont {Li}, \citenamefont {Neven},
  \citenamefont {Platt}, \citenamefont {Smelyanskiy},\ and\ \citenamefont
  {Boixo}}]{villalonga2021efficient}%
  \BibitemOpen
  \bibfield  {author} {\bibinfo {author} {\bibfnamefont {B.}~\bibnamefont
  {Villalonga}}, \bibinfo {author} {\bibfnamefont {M.~Y.}\ \bibnamefont {Niu}},
  \bibinfo {author} {\bibfnamefont {L.}~\bibnamefont {Li}}, \bibinfo {author}
  {\bibfnamefont {H.}~\bibnamefont {Neven}}, \bibinfo {author} {\bibfnamefont
  {J.~C.}\ \bibnamefont {Platt}}, \bibinfo {author} {\bibfnamefont {V.~N.}\
  \bibnamefont {Smelyanskiy}},\ and\ \bibinfo {author} {\bibfnamefont
  {S.}~\bibnamefont {Boixo}},\ }\bibfield  {title} {\bibinfo {title} {Efficient
  approximation of experimental gaussian boson sampling},\ }\href@noop {}
  {\bibfield  {journal} {\bibinfo  {journal} {arXiv preprint
  quant-ph/2109.11525}\ } (\bibinfo {year} {2021})}\BibitemShut {NoStop}%
\bibitem [{foo()}]{footnote}%
  \BibitemOpen
  \href@noop {} {}\bibinfo {note} {Motivated by the evidence of efficient
  tensor network simulation of GBS under high loss, some authors of our work
  also released new results on Gaussian boson sampling simulations using tensor
  networks during the review process of this manuscript. The new work
  significantly reduces the cost and simulates all supremacy experiments while
  obtaining better benchmarking results that can be verified with reasonable
  resources \cite{oh2023tensor}. Efficient simulation using this modified
  tensor network representation under the $N_{\text{out}}\propto \sqrt{N}$
  scaling is rigorously proven.}\BibitemShut {Stop}%
\bibitem [{\citenamefont {Huang}\ \emph {et~al.}(2019)\citenamefont {Huang},
  \citenamefont {Bao},\ and\ \citenamefont {Guo}}]{huang2019simulating}%
  \BibitemOpen
  \bibfield  {author} {\bibinfo {author} {\bibfnamefont {H.-L.}\ \bibnamefont
  {Huang}}, \bibinfo {author} {\bibfnamefont {W.-S.}\ \bibnamefont {Bao}},\
  and\ \bibinfo {author} {\bibfnamefont {C.}~\bibnamefont {Guo}},\ }\bibfield
  {title} {\bibinfo {title} {Simulating the dynamics of single photons in boson
  sampling devices with matrix product states},\ }\href
  {https://doi.org/10.1103/PhysRevA.100.032305} {\bibfield  {journal} {\bibinfo
   {journal} {Phys. Rev. A}\ }\textbf {\bibinfo {volume} {100}},\ \bibinfo
  {pages} {032305} (\bibinfo {year} {2019})}\BibitemShut {NoStop}%
\bibitem [{\citenamefont {Singh}\ \emph {et~al.}(2011)\citenamefont {Singh},
  \citenamefont {Pfeifer},\ and\ \citenamefont {Vidal}}]{singh2011tensor}%
  \BibitemOpen
  \bibfield  {author} {\bibinfo {author} {\bibfnamefont {S.}~\bibnamefont
  {Singh}}, \bibinfo {author} {\bibfnamefont {R.~N.~C.}\ \bibnamefont
  {Pfeifer}},\ and\ \bibinfo {author} {\bibfnamefont {G.}~\bibnamefont
  {Vidal}},\ }\bibfield  {title} {\bibinfo {title} {Tensor network states and
  algorithms in the presence of a global {U(1)} symmetry},\ }\href
  {https://doi.org/10.1103/PhysRevB.83.115125} {\bibfield  {journal} {\bibinfo
  {journal} {Phys. Rev. B}\ }\textbf {\bibinfo {volume} {83}},\ \bibinfo
  {pages} {115125} (\bibinfo {year} {2011})}\BibitemShut {NoStop}%
\bibitem [{\citenamefont {Guo}\ and\ \citenamefont
  {Poletti}(2019)}]{guo2019matrix}%
  \BibitemOpen
  \bibfield  {author} {\bibinfo {author} {\bibfnamefont {C.}~\bibnamefont
  {Guo}}\ and\ \bibinfo {author} {\bibfnamefont {D.}~\bibnamefont {Poletti}},\
  }\bibfield  {title} {\bibinfo {title} {Matrix product states with adaptive
  global symmetries},\ }\href {https://doi.org/10.1103/PhysRevB.100.134304}
  {\bibfield  {journal} {\bibinfo  {journal} {Phys. Rev. B}\ }\textbf {\bibinfo
  {volume} {100}},\ \bibinfo {pages} {134304} (\bibinfo {year}
  {2019})}\BibitemShut {NoStop}%
\bibitem [{\citenamefont {Aizenman}\ \emph {et~al.}(2004)\citenamefont
  {Aizenman}, \citenamefont {Lieb}, \citenamefont {Seiringer}, \citenamefont
  {Solovej},\ and\ \citenamefont {Yngvason}}]{aizenman2004bose}%
  \BibitemOpen
  \bibfield  {author} {\bibinfo {author} {\bibfnamefont {M.}~\bibnamefont
  {Aizenman}}, \bibinfo {author} {\bibfnamefont {E.~H.}\ \bibnamefont {Lieb}},
  \bibinfo {author} {\bibfnamefont {R.}~\bibnamefont {Seiringer}}, \bibinfo
  {author} {\bibfnamefont {J.~P.}\ \bibnamefont {Solovej}},\ and\ \bibinfo
  {author} {\bibfnamefont {J.}~\bibnamefont {Yngvason}},\ }\bibfield  {title}
  {\bibinfo {title} {Bose-einstein quantum phase transition in an optical
  lattice model},\ }\href {https://doi.org/10.1103/PhysRevA.70.023612}
  {\bibfield  {journal} {\bibinfo  {journal} {Phys. Rev. A}\ }\textbf {\bibinfo
  {volume} {70}},\ \bibinfo {pages} {023612} (\bibinfo {year}
  {2004})}\BibitemShut {NoStop}%
\bibitem [{\citenamefont {Alcaraz}\ \emph {et~al.}(1989)\citenamefont
  {Alcaraz}, \citenamefont {Grimm},\ and\ \citenamefont
  {Rittenberg}}]{alcaraz1989}%
  \BibitemOpen
  \bibfield  {author} {\bibinfo {author} {\bibfnamefont {F.}~\bibnamefont
  {Alcaraz}}, \bibinfo {author} {\bibfnamefont {U.}~\bibnamefont {Grimm}},\
  and\ \bibinfo {author} {\bibfnamefont {V.}~\bibnamefont {Rittenberg}},\
  }\bibfield  {title} {\bibinfo {title} {The xxz heisenberg chain, conformal
  invariance and the operator content of c < 1 systems},\ }\href
  {https://doi.org/10.1016/0550-3213(89)90066-7} {\bibfield  {journal}
  {\bibinfo  {journal} {Nuclear Physics B}\ }\textbf {\bibinfo {volume}
  {316}},\ \bibinfo {pages} {735} (\bibinfo {year} {1989})}\BibitemShut
  {NoStop}%
\bibitem [{\citenamefont {Kitagawa}\ \emph {et~al.}(2010)\citenamefont
  {Kitagawa}, \citenamefont {Rudner}, \citenamefont {Berg},\ and\ \citenamefont
  {Demler}}]{kitagawa2010exploring}%
  \BibitemOpen
  \bibfield  {author} {\bibinfo {author} {\bibfnamefont {T.}~\bibnamefont
  {Kitagawa}}, \bibinfo {author} {\bibfnamefont {M.~S.}\ \bibnamefont
  {Rudner}}, \bibinfo {author} {\bibfnamefont {E.}~\bibnamefont {Berg}},\ and\
  \bibinfo {author} {\bibfnamefont {E.}~\bibnamefont {Demler}},\ }\bibfield
  {title} {\bibinfo {title} {Exploring topological phases with quantum walks},\
  }\href {https://doi.org/10.1103/PhysRevA.82.033429} {\bibfield  {journal}
  {\bibinfo  {journal} {Phys. Rev. A}\ }\textbf {\bibinfo {volume} {82}},\
  \bibinfo {pages} {033429} (\bibinfo {year} {2010})}\BibitemShut {NoStop}%
\bibitem [{\citenamefont {Childs}\ \emph {et~al.}(2013)\citenamefont {Childs},
  \citenamefont {Gosset},\ and\ \citenamefont {Webb}}]{childs2013universal}%
  \BibitemOpen
  \bibfield  {author} {\bibinfo {author} {\bibfnamefont {A.~M.}\ \bibnamefont
  {Childs}}, \bibinfo {author} {\bibfnamefont {D.}~\bibnamefont {Gosset}},\
  and\ \bibinfo {author} {\bibfnamefont {Z.}~\bibnamefont {Webb}},\ }\bibfield
  {title} {\bibinfo {title} {Universal computation by multiparticle quantum
  walk},\ }\href {https://doi.org/10.1126/science.1229957} {\bibfield
  {journal} {\bibinfo  {journal} {Science}\ }\textbf {\bibinfo {volume}
  {339}},\ \bibinfo {pages} {791} (\bibinfo {year} {2013})}\BibitemShut
  {NoStop}%
\bibitem [{\citenamefont {Cai}\ \emph {et~al.}(2021)\citenamefont {Cai},
  \citenamefont {Yang}, \citenamefont {Shi}, \citenamefont {Lee}, \citenamefont
  {Andrei},\ and\ \citenamefont {Guan}}]{cai2021multiparticle}%
  \BibitemOpen
  \bibfield  {author} {\bibinfo {author} {\bibfnamefont {X.}~\bibnamefont
  {Cai}}, \bibinfo {author} {\bibfnamefont {H.}~\bibnamefont {Yang}}, \bibinfo
  {author} {\bibfnamefont {H.-L.}\ \bibnamefont {Shi}}, \bibinfo {author}
  {\bibfnamefont {C.}~\bibnamefont {Lee}}, \bibinfo {author} {\bibfnamefont
  {N.}~\bibnamefont {Andrei}},\ and\ \bibinfo {author} {\bibfnamefont {X.-W.}\
  \bibnamefont {Guan}},\ }\bibfield  {title} {\bibinfo {title} {Multiparticle
  quantum walks and fisher information in one-dimensional lattices},\ }\href
  {https://doi.org/10.1103/PhysRevLett.127.100406} {\bibfield  {journal}
  {\bibinfo  {journal} {Phys. Rev. Lett.}\ }\textbf {\bibinfo {volume} {127}},\
  \bibinfo {pages} {100406} (\bibinfo {year} {2021})}\BibitemShut {NoStop}%
\bibitem [{\citenamefont {Schreiber}\ \emph {et~al.}(2012)\citenamefont
  {Schreiber}, \citenamefont {Gábris}, \citenamefont {Rohde}, \citenamefont
  {Laiho}, \citenamefont {Štefaňák}, \citenamefont {Potoček}, \citenamefont
  {Hamilton}, \citenamefont {Jex},\ and\ \citenamefont
  {Silberhorn}}]{schreiber2012a}%
  \BibitemOpen
  \bibfield  {author} {\bibinfo {author} {\bibfnamefont {A.}~\bibnamefont
  {Schreiber}}, \bibinfo {author} {\bibfnamefont {A.}~\bibnamefont {Gábris}},
  \bibinfo {author} {\bibfnamefont {P.~P.}\ \bibnamefont {Rohde}}, \bibinfo
  {author} {\bibfnamefont {K.}~\bibnamefont {Laiho}}, \bibinfo {author}
  {\bibfnamefont {M.}~\bibnamefont {Štefaňák}}, \bibinfo {author}
  {\bibfnamefont {V.}~\bibnamefont {Potoček}}, \bibinfo {author}
  {\bibfnamefont {C.}~\bibnamefont {Hamilton}}, \bibinfo {author}
  {\bibfnamefont {I.}~\bibnamefont {Jex}},\ and\ \bibinfo {author}
  {\bibfnamefont {C.}~\bibnamefont {Silberhorn}},\ }\bibfield  {title}
  {\bibinfo {title} {A 2d quantum walk simulation of two-particle dynamics},\
  }\href {https://doi.org/10.1126/science.1218448} {\bibfield  {journal}
  {\bibinfo  {journal} {Science}\ }\textbf {\bibinfo {volume} {336}},\ \bibinfo
  {pages} {55} (\bibinfo {year} {2012})}\BibitemShut {NoStop}%
\bibitem [{\citenamefont {Agrawal}\ \emph {et~al.}(2022)\citenamefont
  {Agrawal}, \citenamefont {Zabalo}, \citenamefont {Chen}, \citenamefont
  {Wilson}, \citenamefont {Potter}, \citenamefont {Pixley}, \citenamefont
  {Gopalakrishnan},\ and\ \citenamefont {Vasseur}}]{agrawal2022entanglement}%
  \BibitemOpen
  \bibfield  {author} {\bibinfo {author} {\bibfnamefont {U.}~\bibnamefont
  {Agrawal}}, \bibinfo {author} {\bibfnamefont {A.}~\bibnamefont {Zabalo}},
  \bibinfo {author} {\bibfnamefont {K.}~\bibnamefont {Chen}}, \bibinfo {author}
  {\bibfnamefont {J.~H.}\ \bibnamefont {Wilson}}, \bibinfo {author}
  {\bibfnamefont {A.~C.}\ \bibnamefont {Potter}}, \bibinfo {author}
  {\bibfnamefont {J.~H.}\ \bibnamefont {Pixley}}, \bibinfo {author}
  {\bibfnamefont {S.}~\bibnamefont {Gopalakrishnan}},\ and\ \bibinfo {author}
  {\bibfnamefont {R.}~\bibnamefont {Vasseur}},\ }\bibfield  {title} {\bibinfo
  {title} {Entanglement and charge-sharpening transitions in u(1) symmetric
  monitored quantum circuits},\ }\href
  {https://doi.org/10.1103/PhysRevX.12.041002} {\bibfield  {journal} {\bibinfo
  {journal} {Phys. Rev. X}\ }\textbf {\bibinfo {volume} {12}},\ \bibinfo
  {pages} {041002} (\bibinfo {year} {2022})}\BibitemShut {NoStop}%
\bibitem [{\citenamefont {Deshpande}\ \emph {et~al.}(2022)\citenamefont
  {Deshpande} \emph {et~al.}}]{Deshpande2022quantum}%
  \BibitemOpen
  \bibfield  {author} {\bibinfo {author} {\bibfnamefont {A.}~\bibnamefont
  {Deshpande}} \emph {et~al.},\ }\bibfield  {title} {\bibinfo {title} {Quantum
  computational advantage via high-dimensional gaussian boson sampling},\
  }\href {https://doi.org/10.1126/sciadv.abi7894} {\bibfield  {journal}
  {\bibinfo  {journal} {Science Advances}\ }\textbf {\bibinfo {volume} {8}},\
  \bibinfo {pages} {eabi7894} (\bibinfo {year} {2022})}\BibitemShut {NoStop}%
\bibitem [{\citenamefont {Russell}\ \emph {et~al.}(2017)\citenamefont
  {Russell}, \citenamefont {Chakhmakhchyan}, \citenamefont {O’Brien},\ and\
  \citenamefont {Laing}}]{Russell2017direct}%
  \BibitemOpen
  \bibfield  {author} {\bibinfo {author} {\bibfnamefont {N.~J.}\ \bibnamefont
  {Russell}}, \bibinfo {author} {\bibfnamefont {L.}~\bibnamefont
  {Chakhmakhchyan}}, \bibinfo {author} {\bibfnamefont {J.~L.}\ \bibnamefont
  {O’Brien}},\ and\ \bibinfo {author} {\bibfnamefont {A.}~\bibnamefont
  {Laing}},\ }\bibfield  {title} {\bibinfo {title} {Direct dialling of haar
  random unitary matrices},\ }\href {https://doi.org/10.1088/1367-2630/aa60ed}
  {\bibfield  {journal} {\bibinfo  {journal} {New Journal of Physics}\ }\textbf
  {\bibinfo {volume} {19}},\ \bibinfo {pages} {033007} (\bibinfo {year}
  {2017})}\BibitemShut {NoStop}%
\bibitem [{\citenamefont {Nguyen}\ \emph {et~al.}(2022)\citenamefont {Nguyen},
  \citenamefont {Lyakh}, \citenamefont {Dumitrescu}, \citenamefont {Clark},
  \citenamefont {Larkin},\ and\ \citenamefont {McCaskey}}]{nguyen2022tensor}%
  \BibitemOpen
  \bibfield  {author} {\bibinfo {author} {\bibfnamefont {T.}~\bibnamefont
  {Nguyen}}, \bibinfo {author} {\bibfnamefont {D.}~\bibnamefont {Lyakh}},
  \bibinfo {author} {\bibfnamefont {E.}~\bibnamefont {Dumitrescu}}, \bibinfo
  {author} {\bibfnamefont {D.}~\bibnamefont {Clark}}, \bibinfo {author}
  {\bibfnamefont {J.}~\bibnamefont {Larkin}},\ and\ \bibinfo {author}
  {\bibfnamefont {A.}~\bibnamefont {McCaskey}},\ }\bibfield  {title} {\bibinfo
  {title} {Tensor network quantum virtual machine for simulating quantum
  circuits at exascale},\ }\bibfield  {journal} {\bibinfo  {journal} {ACM
  Transactions on Quantum Computing}\ }\textbf {\bibinfo {volume} {4}},\ \href
  {https://doi.org/10.1145/3547334} {10.1145/3547334} (\bibinfo {year}
  {2022})\BibitemShut {NoStop}%
\bibitem [{\citenamefont {Lyakh}\ \emph {et~al.}(2022)\citenamefont {Lyakh},
  \citenamefont {Nguyen}, \citenamefont {Claudino}, \citenamefont
  {Dumitrescu},\ and\ \citenamefont {McCaskey}}]{lyakh2022exatn}%
  \BibitemOpen
  \bibfield  {author} {\bibinfo {author} {\bibfnamefont {D.~I.}\ \bibnamefont
  {Lyakh}}, \bibinfo {author} {\bibfnamefont {T.}~\bibnamefont {Nguyen}},
  \bibinfo {author} {\bibfnamefont {D.}~\bibnamefont {Claudino}}, \bibinfo
  {author} {\bibfnamefont {E.}~\bibnamefont {Dumitrescu}},\ and\ \bibinfo
  {author} {\bibfnamefont {A.~J.}\ \bibnamefont {McCaskey}},\ }\bibfield
  {title} {\bibinfo {title} {Exatn: Scalable gpu-accelerated high-performance
  processing of general tensor networks at exascale},\ }\href@noop {}
  {\bibfield  {journal} {\bibinfo  {journal} {Appl. Math. Stat.}\ }\textbf
  {\bibinfo {volume} {8}} (\bibinfo {year} {2022})}\BibitemShut {NoStop}%
\bibitem [{\citenamefont {Lykov}\ \emph {et~al.}(2021)\citenamefont {Lykov},
  \citenamefont {Chen}, \citenamefont {Chen}, \citenamefont {Keipert},
  \citenamefont {Zhang}, \citenamefont {Gibbs},\ and\ \citenamefont
  {Alexeev}}]{lykov2021performance}%
  \BibitemOpen
  \bibfield  {author} {\bibinfo {author} {\bibfnamefont {D.}~\bibnamefont
  {Lykov}}, \bibinfo {author} {\bibfnamefont {A.}~\bibnamefont {Chen}},
  \bibinfo {author} {\bibfnamefont {H.}~\bibnamefont {Chen}}, \bibinfo {author}
  {\bibfnamefont {K.}~\bibnamefont {Keipert}}, \bibinfo {author} {\bibfnamefont
  {Z.}~\bibnamefont {Zhang}}, \bibinfo {author} {\bibfnamefont
  {T.}~\bibnamefont {Gibbs}},\ and\ \bibinfo {author} {\bibfnamefont
  {Y.}~\bibnamefont {Alexeev}},\ }\bibfield  {title} {\bibinfo {title}
  {Performance evaluation and acceleration of the qtensor quantum circuit
  simulator on gpus},\ }in\ \href {https://doi.org/10.1109/QCS54837.2021.00007}
  {\emph {\bibinfo {booktitle} {2021 IEEE/ACM Second International Workshop on
  Quantum Computing Software (QCS)}}}\ (\bibinfo {year} {2021})\ pp.\ \bibinfo
  {pages} {27--34}\BibitemShut {NoStop}%
\bibitem [{\citenamefont {Lykov}\ \emph {et~al.}(2022)\citenamefont {Lykov},
  \citenamefont {Schutski}, \citenamefont {Galda}, \citenamefont {Vinokur},\
  and\ \citenamefont {Alexeev}}]{lykov2022tensor}%
  \BibitemOpen
  \bibfield  {author} {\bibinfo {author} {\bibfnamefont {D.}~\bibnamefont
  {Lykov}}, \bibinfo {author} {\bibfnamefont {R.}~\bibnamefont {Schutski}},
  \bibinfo {author} {\bibfnamefont {A.}~\bibnamefont {Galda}}, \bibinfo
  {author} {\bibfnamefont {V.}~\bibnamefont {Vinokur}},\ and\ \bibinfo {author}
  {\bibfnamefont {Y.}~\bibnamefont {Alexeev}},\ }\bibfield  {title} {\bibinfo
  {title} {Tensor network quantum simulator with step-dependent
  parallelization},\ }in\ \href {https://doi.org/10.1109/QCE53715.2022.00081}
  {\emph {\bibinfo {booktitle} {2022 IEEE International Conference on Quantum
  Computing and Engineering (QCE)}}}\ (\bibinfo {year} {2022})\ pp.\ \bibinfo
  {pages} {582--593}\BibitemShut {NoStop}%
\bibitem [{\citenamefont {Kerr}\ \emph {et~al.}(2017)\citenamefont {Kerr},
  \citenamefont {Merrill}, \citenamefont {Demouth},\ and\ \citenamefont
  {Tran}}]{kerr}%
  \BibitemOpen
  \bibfield  {author} {\bibinfo {author} {\bibfnamefont {A.}~\bibnamefont
  {Kerr}}, \bibinfo {author} {\bibfnamefont {D.}~\bibnamefont {Merrill}},
  \bibinfo {author} {\bibfnamefont {J.}~\bibnamefont {Demouth}},\ and\ \bibinfo
  {author} {\bibfnamefont {J.}~\bibnamefont {Tran}},\ }\href
  {https://developer.nvidia.com/blog/cutlass-linear-algebra-cuda/} {\bibinfo
  {title} {Cutlass: Fast linear algebra in cuda c++}} (\bibinfo {year}
  {2017})\BibitemShut {NoStop}%
\bibitem [{\citenamefont {Oh}\ \emph {et~al.}(2023{\natexlab{b}})\citenamefont
  {Oh}, \citenamefont {Liu}, \citenamefont {Alexeev}, \citenamefont
  {Fefferman},\ and\ \citenamefont {Jiang}}]{oh2023tensor}%
  \BibitemOpen
  \bibfield  {author} {\bibinfo {author} {\bibfnamefont {C.}~\bibnamefont
  {Oh}}, \bibinfo {author} {\bibfnamefont {M.}~\bibnamefont {Liu}}, \bibinfo
  {author} {\bibfnamefont {Y.}~\bibnamefont {Alexeev}}, \bibinfo {author}
  {\bibfnamefont {B.}~\bibnamefont {Fefferman}},\ and\ \bibinfo {author}
  {\bibfnamefont {L.}~\bibnamefont {Jiang}},\ }\bibfield  {title} {\bibinfo
  {title} {Tensor network algorithm for simulating experimental gaussian boson
  sampling},\ }\href@noop {} {\bibfield  {journal} {\bibinfo  {journal} {arXiv
  preprint quant-ph/2306.03709}\ } (\bibinfo {year}
  {2023}{\natexlab{b}})}\BibitemShut {NoStop}%
\end{thebibliography}%

\end{document}